\documentclass[11pt]{article}

\usepackage{feynmf} \unitlength=1mm 
\usepackage{epsf,epsfig}
\usepackage{cite}
\usepackage{amssymb}
\usepackage[dvips]{color}
\usepackage{amsmath}
\usepackage{amsfonts}
\usepackage{latexsym}
\usepackage{slashed}
\usepackage{mathrsfs}
\usepackage{ifthen}

\let\non\nonumber

\global\newcount\itemno \global\itemno=0

\def\itemaut#1{\global\advance\itemno by1\noindent\item{\the\itemno.}#1}

\newif{\ifeq}           
\eqtrue                 

\newcommand{\be}{\begin{equation}}
\newcommand{\ee}{\end{equation}}
\newcommand{\bes}{\begin{equation*}}
\newcommand{\ees}{\end{equation*}}
\newcommand{\bea}{\begin{eqnarray}}
\newcommand{\eea}{\end{eqnarray}}
\newcommand{\beas}{\begin{eqnarray*}}
\newcommand{\eeas}{\end{eqnarray*}}

\newcommand{\nn}{\nonumber}
\newcommand{\scz}{\setcounter{equation}{0}}

\makeatletter
\renewcommand\section{\@startsection {section}{1}{\z@}%
                                   {-3.5ex \@plus -1ex \@minus -.2ex}
                                   {2.3ex \@plus.2ex}%
                                   {\normalfont\large\bfseries}}
\renewcommand\subsection{\@startsection{subsection}{2}{\z@}%
                                     {-3.25ex\@plus -1ex \@minus -.2ex}%
                                     {1.5ex \@plus .2ex}%
                                     {\normalfont\bfseries}}
\makeatother
\let\non\nonumber

\newcommand{\lp}{\left(}
\newcommand{\rp}{\right)}
\newcommand{\lb}{\left[}

\renewcommand{\(}{\left(}
\renewcommand{\)}{\right)}
\renewcommand{\[}{\left[}
\renewcommand{\]}{\right]}

\newcommand{\pp}[1]{\frac{\p}{\p #1}}

\newcommand{\half}{\frac{1}{2}}

\newcommand{\rt}{{\sqrt 2}}

\renewcommand{\a}{\alpha}

\renewcommand{\d}{\delta}
\newcommand{\g}{\gamma}

\renewcommand{\r}{\rho}
\newcommand{\s}{\sigma}
\renewcommand{\l}{\lambda}
\renewcommand{\lb}{\bar{\lambda}}
\newcommand{\w}{\omega}
\newcommand{\m}{\mu}
\newcommand{\n}{\nu}
\newcommand{\e}{\epsilon}
\newcommand{\eps}{\epsilon}

\renewcommand{\t}{\theta}

\newcommand{\tb}{\bar{\theta}}

\newcommand{\G}{\Gamma}
\newcommand{\T}{\Theta}
\newcommand{\U}{\Upsilon}

\renewcommand{\S}{\Sigma}

\newcommand{\A}{{\cal A}}

\newcommand{\E}{{\cal E}}
\newcommand{\F}{{\cal F}}

\newcommand{\N}{{\cal N}}
\renewcommand{\O}{{\cal O}}
\newcommand{\V}{{\cal V}}

\renewcommand{\L}{{\cal L}}

\newcommand{\J}{{\cal J}}

\newcommand{\W}{{\cal W}}

\newcommand{\IC}{{\mathbb C}}

\newcommand{\IP}{{\mathbb P}}

\newcommand{\IZ}{{\mathbb Z}}

\newcommand{\Tr}{{\rm Tr \,}}
\newcommand{\tr}{{\rm tr \,}}

\def\mod{{\rm mod}}

\newcommand{\lsim}{\,\raise.3ex\hbox{$<$\kern-.75em\lower1ex\hbox{$\sim$}}\,}
\newcommand{\gsim}{\,\raise.3ex\hbox{$>$\kern-.75em\lower1ex\hbox{$\sim$}}\,}

\def\p{\partial}
\def\pb{{\overline \p}}

\newcommand{\tp}{{\theta^+}}
\newcommand{\tbp}{{\bar{\theta}^+}}

\newcommand{\D}{{\rm D}}
\newcommand{\DB}{\overline{\rm D}}
\newcommand{\CD}{{\cal D}}
\newcommand{\CDB}{\overline{\cal D}}

\newcommand{\CDBp}{\overline{\cal D}_{+}}

\newcommand\ZZ[2]{{#1}\mathop{\raisebox{-1ex}{\framebox(7,7){~}}}_{\displaystyle #2}}

\def\NLSM{nonlinear sigma model}

\def\susy{supersymmetry}
\def\susic{supersymmetric}

\def\Ka{K\"{a}hler}

\def\nK{non-K\"{a}hler}
\def\TT{$(2,2)$}
\def\ZT{$(0,2)$}

\def\tpb{\tbp}
\def\rk{{\rm rk}}

\begin{document}
\begin{titlepage}
\begin{flushright}
\today\ \\ 
MIT-CTP 4065 
\end{flushright}
\vskip 1in

\begin{center}
{\Large Orbifold Phases of Heterotic Flux Vacua}

\vskip 0.6in Allan Adams
\vskip 0.3in {\it Center for Theoretical Physics \\ Massachusetts Institute of Technology \\ Cambridge, MA  02139 USA}

\end{center}

\vskip 0.8in

\begin{abstract}\noindent
By studying phase transitions in supersymmetric gauge theories with Green-Schwarz anomaly cancellation, a natural relation is found between sigma models on certain \nK\  manifolds with intrinsic torsion and asymmetric Landau-Ginzburg orbifolds.  In these orbifold limits, a quantum anomaly of the orbifold action is cancelled by discrete phases in the partition function.  These intrinsic torsion phases are derived by blowing down cycles supporting non-trivial $H$-flux in the linear model.  This correspondence extends the CY-LG correspondence to special \nK\ manifolds and provides computational tools with which to study the spectra of associated heterotic flux vacua.
\vskip 0.5in

\end{abstract}

\end{titlepage}

\section{Introduction}\scz

Orbifolds provide simple and computationally tractable descriptions of string propagation on non-trivial spacetimes: by concentrating all the curvature at the orbifold fixed points, the bulk of the theory is free, with all non-trivial interactions determined by the structure of the orbifold singularities\cite{Dixon:1985jw,Dixon:1986jc}.  Finding orbifold limits of smooth manifolds thus allows us to study the spectrum and correlation functions of string theory on the smooth geometry by doing much easier calculations in an exact cft.

Unfortunately, not every smooth geometry can be reliably followed to an exact orbifold limit along a marginal direction.  Conversely, many singular geometries are difficult (if not impossible) to interpret as the singular limit of some smooth geometry.  For example, the introduction of discrete fluxes supported on orbifold singularities may ``freeze'' an orbifold singularity \cite{Vafa:1986wx,Vafa:1994rv,Aspinwall:1995rb,de Boer:2001px}.   The string worldsheet theory on such frozen singularities remains non-singular as a CFT, but does not contain marginal operators whose condensation blows-up the singularity to large volume in string units.  

Finding orbifold limits of smooth geometries supporting non-trivial fluxes would be particularly useful for studying the landscape of flux vacua, where non-trivial fluxes are used to lift moduli.  The possibility of building a purely worldsheet description of a string flux vacuum, however, depends on the duality frame. For example, in type II, stabilizing moduli generically requires turning on NS-NS and R-R fluxes, together with orientifolds and D-branes to cancel tadpoles.  These ingredients make a worldsheet analysis technically challenging, though they may be naturally incorporated into a low-energy effective supergravity analysis (see eg \cite{Kachru:2003aw,Grana:2005jc, Douglas:2006es,Berkovits:2002zk}\ and references therein). In heterotic supergravity, on the other hand, stabilizing moduli requires turning on NS-NS and gauge flux together with curvature and dilaton gradients to cancel tadpoles and satisfy the Bianchi identity \cite{Hull:1986kz,Sen:1986mg,Strominger:1986uh}.  While these ingredients make a supergravity analysis difficult\footnote{Indeed, many years elapsed between the identification of the geometric conditions for unbroken \susy\  \cite{Hull:1986kz,Sen:1986mg,Strominger:1986uh}\ and the construction and explication of the first non-trivial \susic\ solutions\cite{Dasgupta:1999ss,Fu:2005sm,Fu:2006vj,Becker:2006et}.}, they {\em are} amenable to a worldsheet analysis, suggesting that it may be possible to find orbifold limits of some heterotic flux vacua.

For the most elementary flux vacua -- Calabi-Yau compactifications in which the fluxes are trivial -- this limit is realized in the Calabi-Yau--Landau-Ginzburg correspondence \cite{Martinec:1989in,Greene:1988ut,Witten:1993yc}, which allows us to identify special points in the Calabi-Yau moduli space where the worldsheet CFT reduces to an exact Landau-Ginzburg orbifold.  At such LG points, CFT techniques may be applied to compute the spectrum and interactions of the low-energy theory without any supergravity approximation.  The question addressed in this paper is whether there exists a generalization of the CY-LG correspondence to the case of heterotic flux vacua  -- ie, might these compactifications have non-geometric limits governed by exact orbifold CFTs?  
 
If so, what would the blown-down CFT look like?  By definition, any interesting heterotic flux compactification must satisfy the Green-Schwarz anomaly condition non-trivially, 
\be\label{eqBI}
dH = \a' \( \tr R\wedge R - \Tr F\wedge F  \) \neq0,
\ee
so the gauge bundle cannot be identical to the tangent bundle.  In an orbifold limit, then, the orbifold action on the gauge bundle will in general differ from the action on spacetime.  The fact that torsion compactifications often contain cycles with string-scale volume 
(note the $\a'$ above) suggests that the spacetime action may be asymmetric, too.  
Meanwhile, to recover a free orbifold CFT in the blown-down limit, all interactions must vanish in the bulk of the orbifold geometry, so all non-trivial curvatures, including the gauge flux, $F$, the geometric curvature, $R$, and the NS-NS 3-form flux, $H$, must be entirely supported at the orbifold fixed loci, with the anomaly canceling on the singular locus.

The idea of an $H$-flux localized on orbifold singularities first appeared in the study of non-standard modular invariant partition functions for worldsheet orbifolds under name ``discrete torsion'' in \cite{Vafa:1986wx}.  
However, since non-trivial DT generally projects out the moduli which would otherwise resolve the orbifold fixed loci in familiar examples, it has not been possible to directly identify discrete torsion phases in an orbifold's partition function with the blow-down of geometric torsion on its smooth resolution.
Finding an explicit example where blowing down $H$ leads to discrete phases in the 1-loop partition function was part of the motivation of this work.

Some light was shed on these questions by the construction of a linear sigma model for torsion geometries \cite{Adams:2006kb,Adams:2009av} in which the basic object is a 2d Abelian gauge theory with chiral \ZT\ \susy\ (see \cite{Witten:1993yc, Distler:1995mi} for more on the \ZT\ GLSM).  Since left- and right-moving fermions transform in different \ZT\ supermultiplets, they may also transform in inequivalent gauge representations, leading in general to a gauge anomaly.  In the ``large radius'' $r\gg1$ phase, this anomaly takes a simple form in terms of the geometry of $M$, the classical Higgs branch of the gauge theory, and $\V_{M}$, the bundle over $M$ to which the left-movers couple, as
$$
\A \sim  \phi^{*}\lp c_{2}(T_{M}) - c_{2}(\V_{M})\rp,
$$
ie the gauge anomaly is the pullback to the worldsheet of (part of) the spacetime $c_{2}$ anomaly.  Vanishing of the worldsheet gauge anomaly, $\A=0$, is thus necessary for the vanishing of this spacetime anomaly.  Of course, spacetime anomaly cancellation works by virtue of an axion, the NS-NS $B_{\m\n}$, which generates a compensating classical anomaly via the Green-Schwarz mechanism (\ref{eqBI}). This too can be pulled back to the worldsheet by adding to the GLSM a worldsheet axion whose classical gauge anomaly cancels the one-loop quantum anomaly in a worldsheet avatar of the Green-Schwarz mechanism \cite{Adams:2006kb}.  The Higgs branch of the resulting theory is a \nK\ complex fibration $X$ over $M$ with holomorphic vector bundle $\V_{X}\neq T_{X}$ and NS-NS 3-form flux $H$ satisfying the full Bianchi identity.  

More generally, the quantum anomaly of a 2d gauge theory with gauge group $G$ may be cancelled by coupling the vector to an asymmetrically gauged WZW model, as discussed in \cite{Adams:2009av}.  To the degree that the resulting theory admits a geometric description, it can be understood as a fibration of the WZW model over the classical Higgs branch of the gauge theory.  In the special case $G=U(1)^{2}\simeq T^{2}$, this construction reduces precisely to the $T^{2}$-fibrations studied in \cite{Adams:2006kb,Dasgupta:1999ss,Fu:2005sm,Fu:2006vj,Becker:2006et}.  Unlike the known abelian examples, the 4d physics deriving from non-abelian examples may have non-zero generation number.

As we shall see by studying the phase structure of such gauged linear sigma models, blowing down the curves supporting non-trivial $H$-flux {\em can} lead to exact orbifold CFTs.   In essence, we will construct a generalization of the Calabi-Yau---Landau-Ginzburg correspondence \cite{Martinec:1989in,Greene:1988ut,Witten:1993yc}\ to a special class of \nK\ heterotic flux compactifications.  More specifically, we will find linear models which smoothly interpolate between (quasi-) geometric phases governed by \ZT\ non-linear sigma models on \nK\ targets supporting gauge and NS-NS 3-form flux, and phases governed by asymmetric WZW-Landau-Ginzburg orbifolds in which a quantum anomaly in the orbifold action on the LG sector is cancelled by classical ``intrinsic torsion'' phases coming from the classically anomalous asymmetric orbifold action on the partner WZW model\footnote{These LG+WZW orbifolds are pleasingly reminiscent of the WZW orbifolds appearing in \cite{Johnson:1994kv,Berglund:1995dv}.}.
The resulting LG orbifold description provides a tool with which to compute the spectrum of this class of heterotic flux vacua via simple modifications of standard computational techniques \cite{Kachru:1993pg,Distler:1993mk}.  This paper will be limited to the construction of the correspondence; a computation of the massless spectrum in explicit examples is presented in a companion paper, \cite{Adams:2009SPEC}.

We begin in Section 2 with a review of the basic structure of heterotic flux compactifications, including several known examples and a few new examples.  Section 3 gives a brief review of the GLSM for \Ka\ targets.  Section 4 introduces the gauged linear sigma model description of \nK\ heterotic flux vacua.  Section 5 studies the small-radius limit of these models and derives the flux generalization of the CY-LG correspondence.  Section 6 concludes with a discussion of potential new directions.  An Appendix presents an extremely schematic counting of families of flux vacua of the form studied in this paper.

\section{Flux Compactifications in Heterotic Supergravity}\scz

Suppose we want to compactify the heterotic string on a 6-dimensional manifold, $X$, so as to preserve $\N$=1 \susy\ in four dimensions\footnote{The point of requiring unbroken \susy\ is, for the moment, purely practical: the first-order BPS equations are much easier to solve, even implicitly, than the second-order equations of motion.  We will relax the requirement of SUSY shortly; in particular, the main results of this paper do not depend on spacetime supersymmetry, though they will exploit worldsheet supersymmetry extensively.}.  This implies a host of geometric constraints on $X$.  First, for the 4d theory to inherit $\N$=1 \susy\ requires that $X$ admit a nowhere vanishing spinor, $\eta$, such that $\eps_{10}=\eps_{4}\otimes\eta$.  
Since it is nowhere vanishing, there is a connection, $\nabla$, on $X$ which parallel transports $\eta$ to itself, ie according to which $\eta$ is covariantly constant, $\nabla\eta=0$; since the specific choice of spinor breaks the $SO(6)\sim SU(4)$ structure group of $X$ to $SU(3)$, the holonomy of $\nabla$ lies in $SU(3)$.
Meanwhile, since $\eta$ is nowhere-vanishing, $X$ comes equipped with a non-degenerate almost complex structure, $I^{i}_{~j}=\eta^{\dagger}\Gamma^{i}_{~j}\eta$, where $I^{2}=-1$, as well as a holomorphic (3,0)-form $\Omega_{ijk}=\eta^{\dagger}\Gamma_{ijk}\eta$, and a hermitian (1,1)-form $J_{i\bar{j}}=\eta^{\dagger}\Gamma_{i\bar{j}}\eta$.  $J$ determines a metric $g$ on $X$ via $g(V,W)=J(V,IW)$.   The metric then determines a new connection, $\nabla_{g}$.

Preserving \susy\ thus requires $X$ to be an almost complex manifold endowed with a connection $\nabla$ of $SU(3)$ holonomy.   It does {\em not} require that $\nabla$ and $\nabla_{g}$ coincide -- in general, $\nabla$ needn't even be symmetric.  By a theorem of Bismut, however, $\nabla$ and $\nabla_{g}$ differ by a unique anti-symmetric 3-form $H$ such that
$$
\nabla = \nabla_{g} + H.
$$
$H$ is a measure of the intrinsic torsion of the SU(3)-structure.  In the special case $H=0$,  $X$ admits a metric connection of SU(3) holonomy, and is thus Calabi-Yau (see eg \cite{Gauntlett:2003cy} for a more detailed exposition).

Taking $X$ to admit an $SU(3)$-structure does not suffice to ensure 4d $\N$=1; we must also enforce the vanishing of the susy variations of the gravitino, dilettino and gaugino.  Together with the Jacobi identity for the resulting superalgebra, these constraints imply that the complex structure on $X$ is  {\it integrable} and that the full configuration, including the hermitian gauge connection, satisfies the equations,
$$
J^{2}\wedge F=0 ~~~~~~~~
H = i(\pb-\p)J  ~~~~~~~~
d(||\Omega||J\wedge J)=0.
$$
The last equation is the condition that the metric is {\em conformally balanced} \!\footnote{A metric on a complex $n$-fold is called balanced if the associated hermitian form satisfies $d(J^{n-1})=0$.  While weaker than the \Ka\ condition (indeed, the strong Lefschetz theorem does not hold), and much less well studied, this is still an interesting constraint.  For an introduction to balanced manifolds, see \cite{Michelsohn:1982}.}; if $H=0$, then $dJ=0$ and the metric is moreover \Ka.   $H$ is thus an obstruction to $X$ being \Ka.

Finally, we must also impose the 1-loop Green-Schwarz anomaly cancellation condition, 
$$
dH = \a' \lp tr R\wedge R - \tr F\wedge F \rp ,
$$
where $R$ is the curvature of the torsion-laced connection.  This changes the story in several important ways.  First, this equation is nonlinear, so proving the existence of solutions in this ({\em a priori} uncontrolled) one-loop approximation is non-trivial.  Secondly, since the left and right hand sides of this equation scale inhomogenously in the global conformal mode of the metric, any solution to this equation has some moduli fixed to string scale, and will generically contain curvature invariants of order string scale.  

Let's make this last point more explicit\footnote{I thank Piljin Yi for several very illuminating conversations on this point.}.  Consider a shift of the conformal mode of the metric, $ds^{2}\to t^{2} ds^{2}$ ($t$ is the string frame conformal factor, and thus a mixture of the volume of the 6-manifold and the 4d string coupling).  Under this rescaling, the hermitian form and NS-NS 3-form scale as $J\to t^{2} J$ and $H\to t^{2} H$, while the curvature two-form scales as $R\to t^{0}R$.  The Binachi identity thus scales as 
$$ 
t^{2} dH = t^{0} \a'R\wedge R + \dots 
$$
In any non-trivial solution, then, the global conformal mode $t$ is fixed in string units in terms of the quantized fluxes of the solution.  This suggests that the generic non-trivial solution of this equation may not have a strict large radius (small-curvature) limit, ie supergravity perturbation theory appears to have a finite, fixed, expansion parameter, and must be taken with a finite, large, grain of salt.  In general, it does not make sense to work perturbatively in $\alpha'$ (though this may be possible in special cases).   An important caveat here is that $t$ is a mixture of the dilaton, $\phi$, and geometric volume, $vol(X)$; only this mixture of $g_{s}$ and the volume is fixed.  We thus always have a one-paramenter family of solutions labeled by the zero mode of the dilaton.  To lift the dilaton we must go beyond string tree-level (which is well beyond the scope of this paper).

	\subsection{The Canonical Example: $T^{2}\to X\to K3$}

For at least one special class of topologies, $T^{2}$-fibrations over $K3$, it is well-known that solutions for the full set of heterotic BPS equations, including the Bianchi identity, do exist.  This can be argued either by duality\footnote{This model is dual to a well-studied IIB string (often pronounced {\em ``F''}) theory compactification on 
$
K3\times T^{2} / (\Omega (-1)^{F_{L}} {\cal I}_{2} )
$
where $\Omega$ is the worldsheet parity operator, $F_{L}$ the left-moving fermion number operator, and ${\cal I}_{2}\sim\IZ_{2}$ acts by reflection on the legs of the $T^{2}$.  This compactification includes 4 D7-branes and 1 O7-plane at each of the 4 fixed points on $T^{2}/\cal{I}_{2}$, with five-form flux
$
F_{5}=dC_{4} -\half C_{2}\wedge H_{3} +\half B_{2}\wedge F_{3}
$
threading the orientifold.  T-dualizing the $T^{2}$-fibres gives type I on $K3\times T^{2}$ with non-trivial RR flux; S-dualizing gives heterotic SO(32) on $T^{2} \to K3$ with non-trivial gauge and 3-form flux.} \cite{Dasgupta:1999ss} or by direct analysis of the heterotic BPS equations \cite{Fu:2005sm,Fu:2006vj,Becker:2006et}.

In these models, the metric, torsion and holomorphic 3-form take a very specific form: the manifold $X$ is taken to be a holomorphic $T^{2}$-fibration over $K3$ with
\bea
ds^{2}_{X} &=& e^{2u}ds^{2}_{K3} + (d\theta_{1}+\a_{1})^{2} + (d\theta_{2}+\a_{2})^{2} \non \\
\vartheta &=&  (d\theta_{1}+\a_{1}) + i (d\theta_{2}+\a_{2})  \non \\
H &=& \sum_l (d\theta_{l}+\a_{l})\wedge\w_{l}.  \non \\
\Omega_{X} &=& \Omega_{K3}\wedge\vartheta  \non 
\eea
Here, $\w_{l}\in H^{2}(K3,\IZ)$ are the curvatures for the two $S^{1}$-bundles for which $\a_{l}$ are local potentials, $\w_{i}=d\a_{i}$, $\vartheta$ is the globally-defined vertical holomorphic 1-form on the $T^{2}$-fibration, $\Omega_{K3}$ is the holomorphic 2-form on K3, and $u$ is a general function on $K3$, the lone unspecified function in this ansatz.  It is straightforward, if tedious, to check that this ansatz satisfies all the $\N=1$ conditions.  The Bianchi identity then translates into a complicated non-linear partial differential equation for the conformal factor, $u$.

In a lovely piece of analysis \cite{Fu:2005sm}, Fu and Yau proved, 
under mild assumptions, the existence\footnote{There are several important difference between the Fu-Yau and Calabi-Yau results.  First, the Fu-Yau result depends on a specific ansatz for the metric, rather than a set of topological conditions.  Secondly, since the salient low energy lagrangian of string theory is not just Einstein-Hilbert but includes an infinite number of higher-curvature corrections, it is far from obvious that a solution of the tree or 1-loop action extends to a solution of the full equations of motion.  Indeed, even in the case of a CY, it is not true that the ricci-flat metric is a solution to string theory; however, as proved in a pair of beautiful papers \cite{Nemeschansky:1986yx,Gross:1986iv}, any classical solution which is, crucially, \Ka\ may be smoothly connected to an exact solution without lifting moduli.  In the \nK\ case there is no general proof, and in general we should expect higher-curvature corrections to lift these classical solutions for generic values of their moduli.} of a solution to this PDE, and thus, for this ansatz, to the full superstring equations of motion\footnote{As mentioned above, the existence of a solution of the full string equations of motion with this topology can alternately be argued via duality.} 
\!at one loop in $\a'$.  Crucial to their analysis is an integrability condition derived by pushing the Bianchi identity down the fibration and integrating it over the K3, giving a simple integer equation,
$$ 
24-c_{2}(\V_{K3}) =  \sum^{2}_{l=1} N_{l}^{a}N^{b}_{l} C_{ab}  = N_{S}^{2} + N_{A}^{2}, 
$$
where the $N^{a}_{i}$ are the integer classes of the two $S^{1}$ bundles, $C_{ab}$ is the intersection form on the 2-cohomology of K3, and the last equality follows form the supersymmetry condition that the curvature of the $T^{2}$ bundle is the sum of a self-dual (2,0)-form and an anti-self-dual (1,1)-form, 
$$
\w_{1} + \w_{2} = \w^{2,0}_{S}+\w^{1,1}_{A}.
$$  
Had we made the \Ka\ and complex structure of the $T^{2}$ fibres free parameters, they would have appeared in this integrability condition so as to fix one combination of them in terms of the integer data; the solution $R_{i}=l_{s}$ with square complex structure is the simplest solution, but is by no means unique.

\subsubsection{Is SUGRA Self-Consistent?}

The fact that the $T^{2}$ fibres have fixed string-scale radii raises an important question: is the 1-loop in $\a'$ SUGRA analysis used above self consistent?  Let's go back to the scaling analysis of the Bianchi identity reviewed above, where we now scale $ds_{K3}^{2}\to t^{2} ds_{K3}^{2}$.  In the case at hand we can also extract the scaling of each term with the flux integers, $N_{i}$, giving, as forms on $K3$, 
$$
dH\sim t^{2}N^{2}\omega\wedge\omega ~~~~~~ tr R\wedge R\sim N^{4} \omega\wedge\omega.
$$
Equating these terms via the Bianchi identity then sets 
$$
t^{2} \sim N^{2}\a' .
$$ 
At first glance this looks good, since taking $N$ large takes $t$ large, too\footnote{Of course, $N^{2}$ is bounded by 24 in this example, but let's pretend we can take $N$ as large as we like -- which we in fact {\em can}, though at the nominal cost of breaking SUSY.}.  However, what we really need is that the curvature {\em invariants} all remain small, and it is easy to check that they do not -- in particular, the Ricci scalar scales as 
$$
R = g^{\mu\nu}R_{\mu\nu}\sim t^{-2}N^{2}\sim \frac{1}{\a'} .
$$  
To get a sense for where the curvature is getting strong, look back at the metric of the $S^{1}$ fibration.  Each $S^{1}$ is fibred non-trivially around some $\IP^{1}$ in the base $K3$ to give an $S^{3}$, with the $H$-flux lacing this $S^{3}$ related to the degree of the Hopf map by the anomaly equation.  The volume of the base $\IP^{1}$ is part of the defining data of the solution, and may be taken as large as one likes; the fibre radius, however, is fixed to $R=1$ in string units.  The resulting $S^{3}$ thus does not have a round metric, but a squashed metric, with sectional curvatures of order 1 regardless of the radius of the $\IP^{1}$.

The upshot is that, without some miraculous protection from higher-curvature corrections, this solution  does not have a strict large radius (small-curvature) limit, so supergravity perturbation theory will get order one corrections at all orders in the $\a'$ expansion and cannot be trusted.  Happily, in at least some of the cases above, a miracle does occur: if all the 2-form curvatures $\w_{i}$ of the $T^{2}$-fibration are anti-self-dual (1,1)-forms, the compactification actually preserves 4d $\N=2$, ensuring that the moduli space is controlled by a prepotential which is perturbatively 1-loop exact.  Of course, if we study more general $\N=1$ examples, such a miracle cannot be relied upon.  It thus behooves us to find descriptions of these compactifications which do not depend on the $\a'$-expansion.  We will turn to this question in the next section.  First, we introduce the local model.

	\subsection{Local Models}

As with more familiar \Ka\ SU(3)-manifolds, life is considerably easier if we work with local models.  The natural move here is to take the base $K3$ to be non-compact, so let's replace the $K3$ above with $\O_{\IP^{1}}(-2)$, the small resolution of $\IC^{2}/\IZ_{2}$, aka Eguchi-Hanson.  

The geometry of the torsionful system is very easy to follow.  The only interesting element of $H^{2}(\O_{\IP^{1}}(-2))\sim\IZ$ is the hyperplane class of the exceptional $\IP^{1}$, so each $S^{1}$-bundle over $\O_{\IP^{1}}(-2)$ is just a Hopf-fibration over this $\IP^{1}$ with total space $S^{3}$.  The $H$-flux, meanwhile, threads this $S^{3}$.  The relation between the radius of the $S^{1}$, the degree of the Hopf map, and the number of units of $H$-flux is just the balancing of the energy stored in the {\em flux}, which wants to drive the $S^{3}$ large, and the {\em positive curvature} of the $S^{3}$, which wants to drive its volume small.  Meanwhile, the integrability condition is 
$$
2 - c_{2}(\V) =  N^{2}_{S} + N^{2}_{A} .
$$
A particularly simple case is to take the gauge bundle $\V$ to be completely trivial, in which case the gauge flux drops out of the equations entirely.  

Since the metric and the (single) harmonic (1,1) form on $\O_{\IP^{1}}(-2)$ are known in closed form, it is straightforward (if tedious) to expand the BI to give an explicit differential equation for the conformal factor, $u$.  Since the 4d metric is rotationally invariant, $u$ may only depend on the radial coordinate, $\r$, so the BI reduces to a non-linear ODE for the scalar function $u(\r)$.  That a solution exists follows form the results of the compact case; unlike the compact case, however, it is reasonable to hope to find explicit solutions (indeed, while this note was in preparation, such a solution was constructed analytically in \cite{Fu:2008ga}).

~

There are many ways to try to generalize these examples: breaking spacetime supersymmetry; using other bases or fibres; double fibrations; non-geometric bases, etc.  To study these, however, it will be useful to have an $\a'$-exact wordsheet CFT description.  We thus postpone further discussion of such generalizations until we have such tools in hand, turning now to a discussion of the gauged linear sigma model for heterotic flux compactifications.

\section{Review of the \ZT\ Gauged Linear Sigma Model}\scz
\def\NLSM{Non-Linear Sigma Model}
The basic strategy of the Gauged Linear Sigma Model (GLSM) begins with a simple observation: a 2d gauge theory runs, in the UV, to a free field theory, but in the IR to a nonlinear sigma model (NLSM) on the moduli space of the gauge theory, $M$, with the metric on $M$ generated by the dynamics of the gauge theory and the RG flow.  Now suppose you want to study a NLSM on some manifold, $X$, for which you do not have an explicit metric.  If you can build a gauge theory whose higgs branch is the manifold you are after, $M\approx X$, you can just as well study the IR limit of the gauge theory.  Moreover, if you can identify a set of RG-invariants (for example, the chiral ring\footnote{Physically, the chiral rings compute the yukawa couplings (superpotentials) of the effective field theory obtained by compactifying string theory on $M$.  Mathematically, these rings define quantum-corrected versions of certain classical cohomology rings of $M$.} if the model is $\N$=2 \susic), you may compute them either in the strongly-interacting IR NLSM on $X$ (which is hard) or in the weakly-coupled UV Gauge theory (which is easy).  Which is a powerful trick.  Indeed, since its introduction in \cite{Witten:1993yc}, the GLSM has become a basic tool in the study of CY compactifications (and, more generally, \Ka\ manifolds of positive first chern class).  In particular, the GLSM allows a simple proof of the CY-LG correspondence.

It was noted in the original discussion of the GLSM that the possibilities are particularly rich when the system has chiral \ZT\ \susy, corresponding to compactifications of the heterotic string.  This section will give a brief review of the structure of \ZT\ gauged theories emphasizing features we will need below (for a more thorough introduction to the GLSM, there is no more beautiful reference than the original paper, \cite{Witten:1993yc}; for a detailed introduction to \ZT\ models, see \cite{Distler:1995mi}.).  The next section will use these tools to build GLSMs for the \nK\ torsion compactifications discussed in section 2, including local and compact models with and without spacetime supersymmetry.  The subsequent section will study the phase structure of these GLSMs, which will lead to a generalization of the familiar CY-LG correspondence to manifolds with intrinsic torsion and moduli stabilization.  For now, we restrict ourselves to a quick review of \ZT\ gauge linear sigma models without flux.

\subsection{\ZT\ Supersymmetry and Supermultiplets}

The \ZT\ supersymmetry algebra is generated by two superderivatives, $\D_{+}$ and
$\DB_{+}$, two translations $\p_{\pm}$, a rotation, $\cal R$, and a $U(1)$ R-current, $J_{+}$, satisfying,
\bea
\D_{+}^2 =\DB_{+}^2 =0 ~~~~&~&~~ \{\D_{+} , \DB_{+} \}=2i~\!\p_{+}  \nn\\
 \left[ \cal R, \D_{+} \right]=-\D_{+}   ~~&~&~~ \left[ \cal R,\DB_{+} \right] =-\DB_{+}   \nn\\
 \left[ J_{+},\DB_{+} \right]=-\DB_{+}  ~~&~&~~ \left[J_+ , \DB_{+} \right]= +\DB_{+} \nn,
\eea
where  $y^{\pm}=y^{0}\pm y^{1}$.  Many of the models we consider will also feature an additional $U(1)$ flavour symmetry generated by a current $J_{-}$ which counts left-moving fermion number.  

We'll find it useful to represent this algebra in superspace.  Expanding in coordinates $(y^+, y^-,\t^+,\tb^+)$, the supergenerators take the form,
$$
\D_+ =   {\p \over \p \tp}  -i\tpb \p_+
~~~~~~~~~~~~~
\DB_+ = - {\p \over \p \tpb} +i\tp  \p_+  .
$$
Unconstrained superfields are arbitrary functions of $(y^+, y^-,\tp, \tbp)$.

\vspace{.5cm}

In \ZT\ models, there are two inequivalent ``chiral'' multiplets annihilated by $\DB_{+}$, the bosonic {\em chiral} multiplet, which contains a right-moving fermion, and the fermionic {\em fermi} multiplet, which contains a left-moving fermion.  A {\em chiral} multiplet $\Phi$ is a bosonic superfield satisfying $\DB_+\Phi=0$, leading to component expansion,  
%
$$
 \Phi= \phi(y) +\rt\theta^+\psi_+(y) -i\tp\tpb\p_+\phi(y).
$$
The action for a chiral superfield is then
$$
\L_{\Phi} = -{i\over 2}\int \! d^2\t~ \bar{\Phi}\p_-\Phi
~=~
  - |\p\phi|^2 + i{\bar{\psi}}_{+}\p_{-} \psi_{+} 
$$
A {\em fermi} supermultiplet $\G$ is a fermionic superfield satisfying $\DB_{+}\G=0$; in components,
$$
\G = \g_- -\rt\tp F -i\tp\tpb\p_+\g_{-} .
$$
The action for a fermi superfield is then
$$
\L_{\G} = -{1\over 2}\int\ d^2\t ~\bar{\G} \G 
= 
i{\bar{\g}}_{-}  \p_{+} \g_{-} + |F|^2 
$$
To introduce a scalar potential, we can turn on a fermionic superpotential of the form
$$
\L_{J} = {1\over \sqrt{2}} \int\!d\tp~ \G J(\Phi) = \g_{-}\psi_{+i} {\p J \over \p \phi_{i}} + FJ(\phi),
$$
where $\G$ is a fermi superfield and $J(\Phi)$ is a holomorphic function of chiral superfields, $\Phi_{i}$.

\vspace{.5cm}

To construct gauge theories, we need to extend the right-moving (super-)derivatives, $\{\D_+, \DB_+, \p_{+}\}$, to gauge covariant derivatives, $\{\CD_+,\CDB_+,\nabla_{+}\}$, satisfying the algebra
$$ 
{\cal{D}}_+^2 = {\bar{\cal{D}}}_+^2 =0, \qquad  \{\ {\cal{D}}_+,{\bar{\cal{D}}}_+ \}\ =2i \nabla_{+}.
$$
%
This implies $\CD_+ = e^{-V} \D_+ e^{V}$ for some lie-algebra valued scalar superfield  $V$, ie
$$
 {\cal{D}}_+ = \pp{\tp} -i \tbp \nabla_{+}
~~~~~~~~~~~~
 {\bar{\cal{D}}}_+ = -\frac{\partial}{\partial {\bar{\theta}}^+} +i \theta^+  \nabla_{+},
$$
where $\nabla_{+}= \p_{+} + \pp{\tp}\DB_{+}V$.  
We also need to promote the left-moving derivative, $\p_{-}$, to a gauge-covariant derivative on superspace,
$\nabla_{-} = \partial_- +i V_{-}$, where $V_{-}$ is again an unconstrained real superfield. 
Under a gauge transformation with chiral gauge parameter $\DB_{+}B=0$, however, the potentials transform as,
$\delta_B V = i (B-\bar{B})$ and $\delta_B V_{-} = \p_-(B+\bar{B})$.
We may thus fix a Wess-Zumino gauge in which the potentials have component expansion
$$
V = \tp\tpb A_{+}
~~~~~~~~~~~~
V_{-}  = A_- -2i \tp \lb_- -2i \tbp \lambda_- +2 \tp\tpb D,
$$
where $A_{\mu}\to A_{\mu}-\p_{\mu}b$ under the surviving $U(1)$.  In particular, this gives
$$
\nabla_{+} = \partial_+ +iA_{+}.
$$
The on-shell content of the gauge multiplet thus includes a vector, $A_{\mu}$, a complex left-moving gaugino, $\l_{-}$, and an auxiliary scalar, $D$.

Finally, the natural field strength superfield is given, as usual, by the commutator
$$
\U_{-}=[\CDB_{+},\nabla_{-}] = \DB_{+}(\p_{-}V+iV_{-}) = -2\l_{-} + 2i\tp(D-iF_{+-}),
$$
in terms of which the gauge kinetic term takes the form,
\bea
\L_{\Upsilon} &=& {1\over 8e^2}\int\!d^2\t ~\bar{\U}\U \\
&=& \frac{1}{2e^{2}}F_{+-}^2  +\frac{i}{e^{2}} \lb_-\p_{+}\l_-  +\frac{1}{2e^{2}} D^2  . 
\eea 
Since$\U$ is a chiral fermion, we can also add an FI term to the superpotential, giving
$$
\L_{\rm FI} = {1\over 4}\int\!d\tp~ t \U + {\rm h.c.} = -rD+\t F_{+-}
$$ 
where $t=ir+\t$  is the complexified FI parameter.

\vspace{.5cm}

Coupling the vector to chiral matter is now straightforward.  Charged chirals, which transform as $\Phi\stackrel{B}{\to}e^{iB}\Phi$, satisfy the covariant constraint $\CDB_+\Phi=0$, and may again be expanded to give,
$$ 
\Phi = \phi +\rt\theta^+\psi_+ -i\tp\tpb \nabla_{+}\phi.
$$
The corresponding gauge invariant Lagrangian is,
\bea
\L_{\Phi} &=& -{i\over 2}\int \! d^2\t~ \bar{\Phi}\nabla_-\Phi  \\
&=&   -{i\over 2}\int \! d^2\t~ \bar{\Phi}_{0}e^{2QV_{+}} \nabla_-\Phi_{0}   \\
&=& - |\nabla\phi|^2 + i{\bar{\psi}}_{+}\nabla_{-} \psi_{+}   + Q D|\phi|^2 -i Q {\sqrt 2} {\bar{\phi}} \l_{-}\psi_{+} , \non     
\eea
where $\Phi=e^{QV_{+}} \Phi_{0}$ s.t. $\DB_{+}\Phi_{0}=0$.  
Similarly, charged fermi multiplets satisfy the covariant constraint $\CDB_+ \G = 0$, and thus have component expansion,
$$
\G = \g_- -\rt\tp G - i\tp \tpb \nabla_+ \g_-,
$$
and gauge-invariant action,
$$
\L_{\G} = -{1\over 2}\int\ d^2\t
 ~\bar{\G} \G 
= 
i{\bar{\g}}_{-}  \nabla_{+} \g_{-} + |G|^2 .
$$

It is sometimes convenient to work with non-chiral fermi multiplets satisfying $\CDB_{+}\G=\E_{\G}$, where $\E_{\G}(\Phi)$ is some polynomial of chiral superfields, $\CDBp \E=0$.  This is particularly natural in models with accidental \TT\ supersymmetry.  For simplicity, we will mostly work with strictly chiral fermi multiplets, $\E$=0.  However, it is occasionally useful to introduce these more general fields.  In particular, the fact that we have only left-moving gauginos can be extremely constraining, as the only way to lift our left-moving fermions (and thus modify the gauge bundle $\V$ to which they couple) is via a superpotential which gives them mass by pairing them with right-moving fermions -- which changes the geometry (the dimension!) of the moduli space.  Which is a bit heavy-handed.  This can be conveniently avoided by introducing additional uncharged chiral multiplets, $\S_{A}$, and modifying the constraint on the fermi multiplets to take the form,
$$
\CDB_{+}\G_{m}=\S_{A}E^{A}_{m}(\Phi),
$$
where the $E^{A}_{m}(\Phi)$ are polynomials in the $\Phi$ with the same gauge charge as $\G_{m}$.  By suitable choices of $E^{A}_{m}(\Phi)$, linear combinations of the $\G_{m}$ may be lifted by pairing with the $\S_{A}$ without altering the zero mode structure of the charged scalars.  Note that, by introducing charge zero scalars,  this also introduces the possibility of a Coulomb branch, something we have so far avoided.  Note, too, that ensuring the chirality of the superpotential $\int d\t^{+} \G_{m} J^{m}(\Phi)$ requires choosing $E$ and $J$ such that $\CDB_{+}(\G_{m} J^{m}(\Phi))=E_{m}(\Phi)J^{m}(\Phi)=0$.  For the most part we will avoid these subtleties by focusing, purely for simplicity, on models with $\E_{m}=0$.

\subsection{The Geometry of the Higgs Branch}\label{OrbGeo}

A garden-variety \ZT\ GLSM thus includes a gauge group (which we'll take to be abelian for simplicity) $G=U(1)^{s}$ with chiral fieldstrengths $\U_{a=1..s}$, some number $d$ of chiral multiplets $\Phi_{i}$ with charges $Q^{a}_{i}$,  another number $r$ of  fermi multiplets $\G_{m}$ with charges $q^{a}_{m}$, and $s$ complexified FI parameters $t^{a}$, all interacting via the lagrangian density
\be\label{LaG}
L =  - {1\over 2} \int \!d^2 \theta ~\left[ i\bar{\Phi}_{i} \nabla_{-} \Phi_{i} +\bar{\G}_{-m} \G_{-m}   
 - {1\over 4e_{a}^2}\bar{\U}_{-a}\U_{-a}\right] ~ + { 1\over 4}\int d\tp~ t^{a}\U_{-a} + {\rm h.c.} .
\ee
This will be the basic model of interest throughout the rest of this paper.  (We could of course add a superpotential, but let's set it to zero for the moment; we'll add it back in later.)  Integrating out the auxilliary $D$-terms gives,
$$
D_{a} = -e_{a}^{2}(\sum_{i}Q^{a}_{i}|\phi_{i}|^{2} - r^{a}),
$$
leading to the classical scalar potential
$$
U = \sum_{a}e_{a}^{2}(\sum_{i}Q^{a}_{i}|\phi_{i}|^{2} - r^{a})^{2}
$$
Note that the gauge coupling is classically dimensionful, with $e_{a}^{2}\to\infty$ in the IR.

Our main concern is where these theories flow at low energies.  Assuming $r^{a}\neq 0$, minimizing the classical potential forces some of the scalars to aquire non-zero vevs, higgsing the vector.  Quotienting out the space of vevs satisfying $D=0$ by gauge equivalence gives the higgs-branch moduli space, $M=D^{-1}(0)/G$.  Out along $M$, the higgsed scalars and vectors pick up masses which scale as $e^{2}r$; at energies beneath $e^{2}r$, we can integrate out the vector and higgs multiplets to generate an effective action for the remaining massless scalars, which coordinatize the Higgs branch, $M$.  The result is thus a \NLSM\ with target space $M$.

\vspace{.5cm}

It is helpful to see this work in a simple example (we will use this example extensively in later sections). Consider a $U(1)$ gauge theory containing two chiral multiplets $\Phi_{i=1,2}$ of charge $Q_{i}=1$, one chiral $P$ of charge $Q_{P}=-2$, and $r$ fermi multiplets $\G_{m}$ of charge $q_{m}$.  For simplicity, let's set the superpotential to zero.  The D-term scalar potential is
$$
U = e^{2}(|\phi_{1}|^{2}+|\phi_{2}|^{2}-2|p|^{2}-r)^{2}.
$$
To what does this theory flow in the IR?


If $r<0$, minimizing the potential forces the norm of $p$ to take a non-zero vev.  Setting the phase of $p$ to zero by a choice of gauge then entirely removes $p$ from the low-energy dynamics.  This is nothing but the super-higgs effect, with the vector superfield eating the chiral $p$ superfield to pick up a mass $m^{2}\sim e^{2}|r|$.  At energies well below $e^{2}|r|$,  the theory reduces to a free field theory for the surviving massless scalars, $\phi_{1,2}$.  For $r\ll-1$, the theory would appear to flow in the deep IR to the (free) non-linear sigma model on $\IC^{2}$.

That's almost right.  However, since $p$ had charge -2, the vev of $p$ did not completely higgs the $U(1)$ gauge group, but left a $\IZ_{2}$ subgroup unbroken.  This $\IZ_{2}$ acts as $(\phi_{1},\phi_{2})\to(-\phi_{1},-\phi_{2})$.  Since this $\IZ_{2}$ is gauged, we must divide by this surviving $\IZ_{2}$ to get the right result.  Deep in the IR, then, our gauge theory should flow to the $\IC^{2}/\IZ_{2}$ orbifold CFT.

Now consider the case $r>0$.  Minimizing the potential now requires that the $\phi_{i}$ cannot both vanish, but must take vevs satisfying
$$
\sum_{i}|\phi_{i}|^{2}=   2|p|^{2}+r.
$$
The $\phi_{i}$ are thus constrained to live on an $S^{3}$ of radius $\sqrt{r+|p|^{2}}$.  We can again fix gauge by setting the phase of $p$ to zero; this again leaves a $\IZ_{2}$ unbroken.  The space of classical solutions modulo gauge equivalence is thus a cone over $S^{3}/\IZ_{2}$, where the $S^{3}$ has radius $\sqrt{r+|p|^{2}}$ and the $\IZ_{2}$ action is induced by the action on $\phi_{i}$, and is thus free on the $S^{3}$.  For $|p|^{2}\gg r$, this asymptotes to the cone $\IC^{2}/\IZ_{2}$; near $|p|=0$, however, the space remains smooth as the action on the radius-$r$ $S^{3}$ remains free.  This is nothing but the $n=2$ Eguchi-Hanson space, ie the smooth non-compact Calabi-Yau 2-fold which arises as the small resolution of the $\IC^{2}/\IZ_{2}$ orbifold.

Note that our gauge-fixing condition for the phase of $p$ did violence to the complex structure of the space of solutions.  
To make the complex structure of the IR physics more transparent, we can exploit the superspace presentation of the model.  Let's start with the Lagrangian in superfield formalism in eq (\ref{LaG}), where the gauge parameter is a chiral superfield and the gauge symmetry is $\IC^{*}$ rather than $U(1)$.  If we are interested in the geometry of the Higgs branch, we might as well go ahead and integrate out the massive vector multiplet, $V_{\pm}$.  Solving the classical EOM gives,
$$
V_{+}= -\ln\(|\phi_{1}|^{2} +|\phi_{2}|^{2}-2|p|^{2}-r\) ~~~~~~ V_{-}=-i\p_{-} V_{+}
$$
Plugging this back into the action then gives,
$$
S ={1\over 8}\int \!d^{2}x ~ \p_{+}\p_{-} \, t\, \ln\(|\phi_{1}|^{2} +|\phi_{2}|^{2}-2|p|^{2} -r\) + {\rm h.c.}
$$
This is precisely the \Ka\ potential for $\O(-2) \to \IP^{1}$, the small resolution of the $\IC^{2}/\IZ_{2}$ orbifold singularity, in homogenous coordinates; fixing the $\IC^{*}$ gauge invariance can be done by setting, for example, $\phi_{1}=1$, giving
$$
S ={1\over 8}\int \!d^{2}x ~ \p_{+}\p_{-} \, t\, \ln\( 1 +|\phi_{2}|^{2}-2|p|^{2} -r\) + {\rm h.c.}
$$
In general, the resulting effective scalar Lagrangian defines a non-linear sigma model,
$$
\L = g_{ij}(z)~\!\p_{+}z^{i}\p_{-}z^{j} ,
$$ 
where the choice of coordinates, $z$, boils down to the choice of gauge.  It is easy to check in our specific example that the resulting action is precisely that of the non-linear sigma model on the Eguchi-Hanson space, with $g$ the \Ka\ metric following from the \Ka\ potential above, and with $r$ controlling the volume of the small resolution.

A very useful fact we will need below is that the fieldstrength, $F_{+-}$, is the pullback to the worldsheet of a 2-form, $\w$, in spacetime, ie $F=z^{*}\w\in H^{2}(M)$. To see this explicitly, note that,
\bea
F_{+-}&=& \p_{[+}A_{-]}(z) \\
&=&\p_{[+}z^{\bar i}\p_{-]}z^{j} \w_{\bar{i}j}(z)
\eea
The equations of notion for the vector and scalars reduce to the spacetime condition 
$$
d\w=0.
$$
To which 2-form in the target space does the fieldstrength correspond?  In our simple EH case, the answer is extremely simple, since there is only one interesting harmonic 2-form, the Hyperplane class corresponding to the exceptional $\IP^{1}$.  More generally, in theories with multiple $U(1)$'s, the fieldstrengths $F_{a}$ are in 1-to-1 correspondence with the Hyperplane classes $H_{a}$ of the target space, $M$.  
The FI parameters, $t^{a}$, thus determine a 2-form on $M$, this is nothing but the complexified hermetian 2-form of our NLSM, $\J=J+iB=t^{a}H_{a}$.

\vspace{.5cm}

So much for the classical geometry of the moduli space -- what about the fermions?  For the right-movers, unbroken \ZT\ ensures that the they transform  as (the pullbacks of) sections of the tangent bundle of the target space, $T_{X}$.  In particular, the worldsheet kinetic term for the right-moving fermions is the pullback to the worldsheet of the dirac operator on $T_{X}$,
$$
\L_{\psi} =  \psi^{i}_{+}(g_{ij}\p_{-}+(\p_{-}z^{k})\Gamma_{ijk})\psi^{j}_{+} .
$$ 
In the linear model, the fact that $\Gamma_{ijk}=\p_{k}g_{ij}$ correspond to the connection on $T_{X}$ follows from the fact that the scalars and their right-moving superpartners carry identical gauge charges, and thus couple to the same low-energy auxiliary vector $A(\phi,p)$.

For the left-movers, \ZT\ requires that the kinetic term is the pullback of a Dirac operator on a holomorphic bundle $\V_{X}$, with $\rk{\V_{X}}$ given by the number of massless left-movers,
$$
\L_{\l} =  \lb_{-m} (h_{mn}\p_{+} + (\p_{+}z^{i}) A_{imn}) \l_{-n}
$$
This bundle does not, however, need to have anything to do with the tangent bundle -- indeed, it needn't even be the same dimension.  The choice of bundle is thus part of the defining data of the NLSM.  This choice is fixed in the linear model by specifying the number of massless\footnote{Since there are no right-moving gauginos, all left-moving fermions in the linear model remain exactly massless until additional non-gauge interactions, such as a superpotential, are turned on.} left-moving fermions and their gauge charges.

Thus, in our gauged linear sigma model, the charges of the chiral multiplets, $Q_{i}$, determine the target space $M$, the charges of the fermi multiplets, $q_{m}$, determine the holomorphic vector bundle $\V_{M}$, and the FI parameters determine the moduli of the complexified hermetian form, $\J=t^{a}H_{a}$.

\subsection{Quantum Anomalies}

Classically, at energies well below the Higgs mass, $E\ll e^{2}r$, our gauge theory reduces to a non-linear sigma model on the Higgs branch of the moduli space.  What happens in the quantum theory?  As usual, $\N=2$  non-renormalization theorems protect the superpotential (and thus scalar potential) from perturbative renormalization beyond 1-loop, so all we need to worry about are 1-loop and non-perturbative corrections.  As explained in \cite{Silverstein:1995re,Beasley:2003fx}, gauge theory instantons do not lift the perturbative moduli space, so we only need to worry about 1-loop effects.  

Two one-loop diagrams have the potential to bite us.  First, scalar loops generate a radiative tadpole for the D-term,
$$
\begin{fmffile}{fmfDterm}
\parbox{20mm}{
\begin{fmfgraph*}(45,30)
	\fmfleft{i}  
	\fmfright{o} 
	\fmf{phantom,tension=7}{i,z1}
	\fmf{dots}{z2,o}
	\fmf{dashes,right, tension=1/3}{z1,z2,z1}
\end{fmfgraph*}}
\hspace{3cm}
\propto
~~~ \lp\sum_{i}Q^{a}_{i}\rp \ln{\frac{\mu}{\Lambda}}
\end{fmffile}
$$
corresponding to a 1-loop renormalization of the superpotential, $\int d\t^{+}\Upsilon_{a} t^{a} = D_{a}r^{a}$+...  This can be folded into a 1-loop renormalization of the FI coupling,
$$
t^{a}_{eff}(\mu) = t^{a} - \lp \sum_{i}Q^{a}_{i} \rp ~\! \ln \lp\frac{\mu}{\Lambda}\rp
$$
Since the FI parameters $r^{a}$ measure the volume of the various non-trivial cycles in the target space, with $\J=t^{a}H_{a}$, this one-loop renormalization of $r^{a}$ drives the corresponding cycle either to zero volume (if $\sum Q^{a}_{i}>0$) or infinite volume (if $\sum Q^{a}_{i}<0$) in the IR.  In either case, the FI parameter represents a non-marginal mode of the IR theory.  If, and only if, $\sum Q^{a}_{i}=0$, does $r^{a}$ remains exactly marginal.  Since a NLSM flows to zero (infinite) volume if the target space is positively (negatively) Ricci-curved, the running of the linear model suggests that the Ricci curvature of the target space is proportional to $\sum Q^{a}_{i}$.  We will see this explicitly below.

Another possibility is a 1-loop chiral anomaly.  Since our left- and right-moving fermions transform in independent gauge multiplets, it is possible for their zero-modes to generate a gauge anomaly.  In 2d, the anomaly is the ``diangle'' diagram,
$$
\begin{fmffile}{fmfAnomaly}
\parbox{20mm}{
\begin{fmfgraph*}(45,30)
	\fmfleft{i} \fmfv{label=$A^{a}_{\mu}$}{i} 
	\fmfright{o} \fmflabel{$A^{b}_{\mu}$}{o}
	\fmf{photon}{i,z1}\fmflabel{$Q^{a}_{I}$}{z1}
	\fmf{fermion,label=$\psi_{+},\g_{-}$,right,tension=1/3}{z1,z2}
	\fmf{fermion,right,tension=1/3}{z2,z1}
	\fmf{photon}{z2,o}\fmflabel{$Q^{b}_{I}$}{z2}
\end{fmfgraph*}}
\hspace{3cm}
\propto
~~~ \sum_{+}Q^{a}_{i}Q^{b}_{i}   - \sum_{-}q^{a}_{m}q^{b}_{m} ~~ \equiv ~\A^{ab}
\end{fmffile}
$$
Under a gauge transformation with gauge parameter $\a$, the measure thus transforms as,
\be\label{ANOM}
\int\!\CD\Phi ~ \stackrel{\a}{\rightarrow} \int\!\CD\Phi ~  e^{\int\!\A^{ab}\a_{a} F_{b}} ~,
\ee
where  $\int\! F$ is the 2d axial term analogous to $\int\! F\wedge F$ in 4d.   Now, as we have seen above, the fieldstrengths $F_{a}$ are the pullback to the worldsheet of hyperplane classes of the target space, $F_{a} = z^{*}H_{a}$, with $H_{a}\in H^{2}(M)$.  The anomaly matrix $\A^{ab}$, which is by construction a symmetric bilinear form on the space of gauge fieldstrengths, is thus the pullback to the worldsheet of a symmetric bilinear form on $H^{2}(M)$,
$$
\A = \A^{ab}H_{a}\wedge H_{b}\in H^{4}(M).
$$
Which 4-form does the anomaly represent?  A standard analysis (see eg \cite{Distler:1995mi}) gives,
$$
\A = c_{2}(T_{M})-c_{2}(\V_{M}).
$$
If we want to use the semi-classical analysis of our GLSM, and in particular of the NLSM to which we believe it flows in the deep IR, we need to deal with this anomaly.  Traditionally, the anomaly is used as a constraint on the valid choices of $Q_{i}$ and $q_{m}$ -- ie, we restrict ourselves to charge assignments which ensure $c_{2}(T_{M})=c_{2}(\V_{M})$.  This ensures that $M$ admits a \Ka\ metric with $H=0$, and is thus topologically a Calabi-Yau\footnote{In general, the RG will generate not the CY metric with torsion-free SU(3) connection, but a small, massive, deformation thereof; if we are not at standard embedding, this will generally include some non-vanishing, but topologically trivial, $dH\neq0$}.   However, as we will explore in the next section, this is not our only option.

Importantly, the gauge current is not the only current subject to a 1-loop anomaly.  For example, in simple models the right-moving R-current couples with unit charge to the right-moving fermions $\psi_{+i}$ in the chiral multiplets, but not to the left-moving fermions or gauginos, and thus has 1-loop anomaly proportional to 
$$
\A^{a}_{R}= \sum_{i} Q^{a}_{i}.
$$
Requiring that the R-current is non-anomalous forces $\sum_{i} Q^{a}_{i}=0$ for each gauge group.  It is well known that non-anomaly of the R-current of an NLSM is equivalent to the target space being Calabi-Yau.  We can see this more directly.  Since $F_{a}=z^{*}H_{a}$, the anomaly coefficient $\A^{a}_{R}$ defines a 2-form on the target space given by $\A_{R}=\A_{R}^{a}H_{a}$; it is a short exercise to show that $\A_{R}$ computes $c_{1}(T_{M})$.  This fits the running of the FI parameters found above.

A similar analysis applies to the left-moving $U(1)_{L}$ current which couples with unit charge to left-moving fermions in fermi multiplets, $\l_{m}$; the one-loop anomaly of the $U(1)_{L}$-current is thus proportional to
$$
\A^{a}_{L}= \sum_{m} q^{a}_{m}.
$$
This again maps to a 2-form on $M$.  Now, however, it must be a characteristic 2-form of the vector bundle, $\V_{M}$, to which the left-moveers couple.  Analogous to the right-moving case, this 2-form is the first chern class, $\A_{L}\sim z^{*}c_{1}(\V_{M})$.  Note, however, that this $U(1)_{L}$ is not in the \ZT\ superconformal group; as such, a non-vanishing anomaly, $\A_{L}\neq 0$, is not an obstruction to our gauge theory flowing to a non-trivial CFT.  However, the heterotic GSO projection does require a left-moving fermion operator, $\IZ_{2}$; to build a good string theory, we thus need $\A_{L}=0~\mod~2$.  This anomaly will play an important role in the computation of the spectrum.

One last note.  Manifest in all our models is the \ZT\ \susy\ of the worldsheet.  Where is {\em spacetime} supersymmetry used, or at least hardwired in?  The answer is that it is not.  All we need for our analysis is worldsheet $\N=2$.  Spacetime supersymmetry, as usual in RNS formalism, arises as an accidental symmetry of the physical spectrum after GSO projection.  There is, however, a simple test for spacetime \susy: for all states in the physical spectrum to pair with degenerate states of opposite spacetime statistics, the $R$-charges of the worldsheet fields must be integer quantized.   Thus, while generic examples of the theories we study will not be spacetime \susic, checking whether a given example is or is not \susic\ is relatively straightforward.

\newcommand{\vt}{\vartheta}
\section{Flux and the Worldsheet Green-Schwarz Mechanism}\scz  

Suppose you are handed a \ZT\ GLSM for some Calabi-Yau $M$ decorated with a bundle $\V_{M}$ such that the gauge anomaly, proportional to $\A=c_{2}(T_{M})-c_{2}(\V_{M})$, is non-vanishing.  In our review of the \ZT\ GLSM, we would have simply ruled out this model.  

There is, however, another option.  Recall that in 10d supergravity, the $c_{2}$ anomaly arises from the mixed gauge and gravitational anomaly -- more precisely, the equation 
$$
dH=tr R\wedge R - Tr F\wedge F
$$
is the condition that the classical SUGRA action, which includes a chern-simons term of the form $B\wedge F^{4}$, has a {\em classical gauge anomaly} which precisely cancels the {\em quantum anomaly} of the measure for the spacetime fermions.  This is the famed Green-Schwarz anomaly cancellation mechanism, with $B_{\mu\nu}$ playing the role of a gauge-charged 2-form axion whose classical lagrangian is gauge-variant.

Let's apply the same mechanism to our worldsheet gauge anomaly.  Rather than requiring the anomaly to cancel (which locked us into the CY closet), let's introduce an axion, $\vt$, whose classical action is gauge-{\em variant} precisely so as to cancel the quantum anomaly.  Looking back at the form of the anomaly in (\ref{ANOM}), we see that the gauge-variation of the classical axion action must be,
$$
\delta_{\a}\L_{\vt} = -\A^{ab}\a_{a}F_{b}.
$$
It's easy to construct a suitable action.  Let $\vt_{l}$ be a doublet of real periodic scalars, $\vt_{l}\sim\vt_{l} +2\pi$, (we need a doublet to fill out a good representation of \ZT\ \susy) with gauge transformation
$$
\vt_{l} \stackrel{\a_{a}}{\rightarrow} \vt_{l} + N_{l}^{a}\a_{a}
$$
and action
$$
\L_{\vt} = R^{2}_{l}(\p\vt_{l} -N_{l}^{a}A_{a})^{2} - M^{lb} ~\! \vt_{l} F_{b}.
$$
While the kinetic terms are gauge-invariant, the axial coupling is not, giving,
$$
\L_{\vt} \stackrel{\a_{a}}{\longrightarrow} \L_{\vt} -M^{lb}N_{l}^{a} ~\! \a_{a} F_{b}.
$$
If we now arrange the charges $M^{lb}$ and $N^{a}_{l}$ such that,
$$
\sum_{l} M^{lb}N_{l}^{a} = \A^{ab},
$$
then the classical variation of the action precisely cancels the quantum anomaly of the measure in a worldsheet avatar of the Green-Schwarz mechanism.

To preserve worldsheet \susy, we must also add superpartners for the axions. 
As a guide, note that $\vt$ is essentially a doublet of dynamical theta angles.
Since the original theta angle comes from a superpotential term,
$$
\L_{\t} = \int \! d\tp ~t^{a}\Upsilon_{-a} = \dots r^{a}D_{a}+\t^{a}F_{a},
$$
the axions must live in a chiral multiplet,
$$
\Theta = \vartheta + i\rt \tp \chi_{+},
$$
where $\vt=\vt_{1}+i\vt_{2}$, and with $\chi$ gauge invariant. A \susy\ transformation of the axion action then leads to the fermionic action,
$$
\L_{\chi} = 2iR^{2}_{l}~\!\bar{\chi}_{+l}\p_{-}\chi_{+l} + (2R^{2}_{l}N^{a}_{l}-M^{a}_{l})\frac{i}{\sqrt{2}}\chi_{+l}\lambda_{-a}
$$
Since the gauginos $\l_{-a}$ transform into $(D_{a}+2iA_{+-a})$, this action is not, in general, \ZT\-invariant: while the $A_{+-}$ term does cancel against the variations of the $\vt_{l}$ kinetic term and the axial superpotential, the $D$ term has nothing against which to cancel.  Preserving \ZT\ \susy\ thus requires that the last term is identically zero -- ie, that the \Ka\ moduli, $k_{l}$, of the $T^{2}$-fibres are fixed in terms of the flux integers $N_{l}$ and $M_{l}$ as,
$$
R^{2}_{l} = \frac{M^{a}_{l}}{2N^{a}_{l}}.
$$
It would be interesting to find models in which this constraint was relaxed, so that $M$ and $N$ could point in independent directions, while still preserving \ZT\ \susy. 
We can certainly take the complex structure of the $T^{2}$ away from the rectangular choice used above; preserving \ZT\ then lifts a non-trivial mixture of the \Ka\ and complex moduli.\footnote{It is intriguing to wonder whether this constraint on the \Ka\ and complex structure moduli is reproduced, in the SUGRA approximation, by the superpotential $W = \int H\wedge\Omega$.
This superpotential has been motivated in numerous ways, but never derived directly from a heterotic worldsheet argument; sharpening this connection, if possible, would be extremely interesting.}  This does not seem to suffice.  For now we focus on models of this form.

\subsection{The Geometry of the Green-Schwarz Higgs Branch}
In canceling the anomaly by introducing an axion, we have altered the theory in several ways.  Perhaps most obvious is the fact that, by explicitly higgsing the vector, the axial term has completely lifted any potential coulomb branches for the anomalous vector. 
More dramatic is what the axions do to the surviving Higgs branch of the moduli space.  In particular, adding two real scalars should increase the real dimension of the Higgs branch by 2.  So: what is the geometry of the Higgs branch?

Repeating the analysis in the previous section for our non-anomalous axionic theory gives a target space metric of the form,
$$
ds_{X}^{2} = e^{u(z)} ds^{2}_{M} + R^{2}_{l}(d\vt_{l}+N^{a}_{l}\tilde{A}_{a}(z))^{2},
$$
where $u$ is a smooth function on the higgs branch of our original gauge theory, $M$, and the $\tilde{A}_{a}$ are simple deformations of the one-form potentials for the hyperplane classes $H_{a}=dA_{a}$ on $M$, as explained in \cite{Adams:2006kb}.  The geometry of the higgs branch is thus a $T^{2}$-fibration over $M$ with curvatures for the two $S^{1}$ bundles given by the integer classes $N^{a}_{l}H_{a}\in H^{2}(M)$.  Unbroken \ZT\ \susy\ ensures that this is a holomorphic fibration.  We'll call the total space $X$, with $\pi$ the projection to $M$, ie $T^{2}\to Y\stackrel{\pi}{\rightarrow}M$.  Since the left-movers do not interact with the axions, the vector bundle they define over $X$ is simply the pullback to $X$ of the holomorphic bundle over $M$, ie $\V_{X}=\pi^{*}\V_{M}$, which is automatically holomorphic.

This is not the end of the story.  In the presence of the axial coupling, integrating out the vector generates {\em antisymmetric} contributions to the scalar kinetic terms of the form,
$$
\L_{eff} 
= \eps^{\mu\nu} \p_{\mu}x^{i} \p_{\nu} x^{j} B_{ij}(x) + ...
$$
where the $x$ are coordinates on the total space, $X$.  
In fact, such a term is already present in non-anomalous CY compactifications -- it comes from the {\em constant} theta-angles, $\theta^{a}$, in the original gauge theory, corresponding to a constant, ungauged $\vt$.  This leads to a globally well-defined $B$-field of the form, $B = \theta^{a} H_{a}$, and thus $H=dB=0$.  Here, $d\vt$ is gauge-variant and thus not globally well-defined; the full globally well-defined object appearing in the place of $dB$ is in fact,
$$
H = M^{la}(d\vt_{l}+N^{b}_{l}A_{b})\wedge F_{a},
$$
as worked out in detail\footnote{In deriving this result, it is important to recall that susy transformations square not to zero but to a gauge transformation in Wess Zumino gauge.  Normally this is merely a formal annoyance; when there is an anomaly, however, preserving \susy\ requires including the anomalous variation of the action under the gauge transformation used to re-fix WZ gauge.  To avoid these subtleties, it is actually considerably easier to undo WZ gauge and work with all auxiliary fields and gauge symmetries manifest.} in \cite{Adams:2006kb}.  While gauge-invariant, and thus globally well-defined, this 3-form flux is not closed, but rather satisfies,
$$
dH = M^{la}N^{b}_{l}~\!F_{a}\wedge F_{b} = \A^{ab}~\! F_{a}\wedge F_{b} = c_{2}(T_{X}) - c_{2}(\V_{X}),
$$
where the second equality follows from anomaly cancellation condition above.   At low energies, then, our GLSM flows to an $\N$=2 NLSM on the fibration $T^{2}\to X\to M$ decorated with a holomorphic bundle pulled back from $M$, $\V_{X}=\pi^{*}\V_{M}$, and supporting non-trivial 3-form flux with $dH\neq0$ satisfying the Bianchi identity.

It's useful to note a consistency check on this geometry.  First, the spacetime potential energy (ie the scalar terms in the 10d $\N$=1 supergravity) includes a term of the form $|H|^{2}$, so the 3-form $H$ had better be globally well defined.  
Our construction began, of course, by setting $[c_{2}(T_{M}) - c_{2}(\V_{M})]\neq 0$, and thus $[dH]\neq0\in H^{4}(M)$.  However, in canceling the resulting anomaly, we lifted $M$ to a fibration $X$ over $M$ on which $H$ {\em is} globally well defined, with $[dH]_{X}=[c_{2}(T_{X})]-[c_{2}(\V_{X})]=0\in H^{4}(X)$.  More precisely, $H$ is a sum of two terms, one proportional to $d\vt_{l}$ and one proportional to $N^{b}_{l}A_{b}$.  Neither is globally well-defined.  The first is naturally interpreted as $dB$, and the second as the chern-simons contribution $\w_{3}(T)-\w_{3}(\V)$. Again, neither is  globally well-defined.  Their sum, however, is proportional to $d\vt_{l}+N^{b}_{l}A_{b}$, which is globally well-defined -- it is the vertical 1-form along the $T^{2}$-fibration.  Thus, on $X$, $H$ is globally well defined, with $c_{2}(T_{X}) - c_{2}(\V_{X})$  cohomologically trivial.

Of course, this is just the geometry as seen by the 1-loop effective action.  As discussed in the section on heterotic flux compactifications, there is no reason to expect such a 1-loop description to be self-consistent.  The virtue of the GLSM is that it is defined without committing to any geometric approximation or one-loop approximation.  By finding phases in which the gauge theory, not the NLSM, simplifies, we will be able to construct the massless spectrum of the string theory without making any supergravity approximations.

	\subsection{Non-Abelian GLSMs and Gauged WZW Models}
The models above can be illuminated and generalized  by replacing the axion multiplet with a gauged WZW model \cite{Adams:2009av}.  Consider a WZW model with target group $G$ and lagrangian,
$$
S = -{k \over 4\pi} \int_{\Sigma} \tr\!\[g^{-1} \p_+ g g^{-1} \p_- g\]  -{ik \over 12 \pi} \int_V   \tr\!\[(g^{-1} \p_i g)(g^{-1} \p_j g)(g^{-1} \p_k g)\] \e^{ijk}
$$
where $g$ is a $G$-valued scalar and $\Sigma=\p V$.  We can gauge this WZW model by either a right-action, $g\to g h$, or a left-action, $g\to hg$.  If we call the vectors gauging the left- and right-actions $A^{L}$ and $A^{R}$, respectively, the coupling can be written as,
$$
\delta S = {ik\over 2\pi} \int_{\Sigma} \tr\!\[g^{-1} \p_+ g A^{R}_- - A^{L}_+ \p_- g g^{-1} + iA^{R}_- g^{-1} A^{L}_+ g + {i \over 2}(A^{L}_+ A^{L}_-+A^{R}_+A^{R}_-)\]
$$
Notably, the resulting classical action is {\em not} gauge-invariant.  Under a gauge variation with gauge parameters $\a_{L}$ and $\a_{R}$, the action varies as,
$$
\d S = {k\over 4\pi}\(\tr\[\a_{R}F^{R}_{+-}\]-\tr\[\a_{L}F^{L}_{+-}\]\)\,,
$$
where $F^{L,R}_{\m\n}$ is the fieldstrength of $A^{L,R}_{\m}$.  To make the action gauge invariant we must choose the diagonal gauging with $A^{R}=A^{L}$.

Of course, in our flux models, we do not want a gauge-invariant classical action. Rather, we want the classical anomaly of the action to cancel the quantum anomaly of the gauge theory -- and indeed the classical anomaly of our gauged WZW model is of just the right form to cancel the chiral anomaly (including its supersymmetric completion), at least if we choose the gauging and level appropriately (see \cite{Adams:2009av}\ for a detailed discussion).

As an example, consider the $U(1)\times U(1)$ WZW model left-gauged by the $U(1)$ action,
$$
(g_{1},g_{2}) \stackrel{\a}{\rightarrow} (e^{i\a N_{1}}g_{1},e^{i\a N_{2}}g_{2}).
$$
Using coordinates $g=(e^{i\vt_{1}},e^{i\vt_{2}})$, the lagrangian for this simple example takes the form,
$$
\L=\frac{k}{4 \pi} ( \partial_+ \theta_l \partial_- \theta_l -  2 N_l A_+ \partial_-\theta_l +  (N_1^2 +N_2^2) A_+ A_-) .
$$
If we identify $A^{L}$ with the vector of the GLSM and set $R^{2}_{l}=k$, this is precisely the form of the axion action in the torsion models above.
 
The upshot is that the torsion linear models discussed above can be reconsidered as gauged LSM-WZW hybrids in which we gauge a non-anomalous current built out of both LSM and WZW fields, with the classical anomaly of the asymmetrically-gauged WZW model canceling the quantum anomaly of the GLSM.  Notably, in this description, there is nothing special about abelian gauge groups -- this construction works just as well with general non-abelian gauge groups and WZW-models on non-abelian targets (or more generally cosets), as discussed in \cite{Adams:2009av}, and indeed we can cancel the anomaly in any GLSM by coupling in a suitable WZW model, leading to new flux compactifications.  The crucial feature is the asymmetric gauging of the WZW model, an immediate generalization of the axial coupling of the torsion models above.

As in the $U(1)$ cases, integrating out the higgsed vector again generates an effective NLSM whose target is a fibration over the classical higgs branch of the GLSM (eg $K3$).  When $G$ is non-abelian, however, the fiber is not a $T^{2}$ but rather an intrinsically non-geometric WZW model on $G$ at low level, $k\sim\O(1)$.  The fact that we are at low level means a NLSM or SUGRA description has no reason to be reliable, as we expect from the structure of the Bianchi identity.  In general, these are {\em non-geometric} flux compactifications.

This WZW presentation solves a minor mystery we've elided in the above.  As mentioned in Section 2, some of these models actually respect spacetime $\N$=2 \susy.  This implies that the worldsheet theories are secretly (0,4), not just (0,2).  But (0,4) \susy\ requires the target space of an NLSM to be $4n$ real dimensional -- and yet our target space would appear to be a complex 3-fold $T^{2}$-fibration over a 2-fold base.   The point is that the theory is not a weakly coupled NLSM on a 3-fold.  More explicitly, recall the beautiful fact that  the $c=1$ CFT corresponding to the NLSM on $S^{1}$ at self-dual radius is isomorphic to the $SU(2)$ WZW model at level $k=1$.  Since the radius of each axionic $S^{1}$ in the models above are also fixed to self-dual radius, the $T^{2}$'s above may be replaced by $SU(2)\times U(1)$ WZW models left-gauged by the vector of the GLSM.  The full $(0,4)$ \susy\ can then be linearly realized on the 4n-dimensional fibration $SU(2)\times U(1)\to K3$.  Of course, this ``4-fold'' is strongly curved, with the corresponding cft having central charge $\hat{c}=3$, not 4 (the ``$4n$-dimensions'' condition arises in the large-radius limit, where engineering dimensions are the correct dimensions, and where these flux vacua very definitely do not live).  The advantage of the 3-fold presentation is that it gets the central charge classically right; the advantage of the 4-fold description is that it allows all the supersymmetry to be linearly realized.  In general, it is just an abstract cft.

	\subsection{Some Examples}

At this point, it is useful to look at some examples.  We'll begin with the non-compact example we introduced in the last section, ie the Eguchi-Hanson model.  We'll then look at some compact models, including models without \susy.

		\subsubsection{Local Models}
		
One particularly simple model is a non-compact fibration over the small resolution of the $A_{1}$ singularity, which is a non-compact K3.  The linear model describing this geometry contains a single $U(1)$ vector with FI parameter $r$, three chiral multiplets with charges $\{1,1,-2\}$, two fermi multiplets of charge $\{2,1\}$, and a torsion multiplet with charge $N_{1}=N_{2}=1$.  These charges have been chosen to ensure the cancellation of the one-loop gauge anomaly against the gauge variation of the classical action for the torsion multiplet.

Let's check that this is the 
model we're looking for.  For $r\gg1$, the 1-loop moduli space is a $T^{2}$-fibration over $\O(-2)\to\IP^{1}$, a non-compact $K3$, with the $T^{2}$ fibred as the unit sub-bundle of $\O(1)\oplus\O(1)$ over the $\IP^{1}$.  The geometry is further decorated with a non-trivial gauge bundle, $\V=\O(2)\oplus\O(1)$, and non-vanishing 3-form flux, $H=d\t\wedge J$, where $d\t$ is the vertical (1,0)-form along the elliptic fibre and $J$ is the \Ka\ class of the $\IP^{1}$.  Which is just what we wanted at $r\gg1$.  Generalizing this is straightforward.

		\subsubsection{Compact $\N$=2 Models}
		
The basic change in going to a compact model is the introduction of a superpotential.  Consider the $T^{2}$ fibration over the sextic in $W\IP_{1,1,2,2}$, with non-trivial $H$ supported entirely on the exceptional divisor.  The linear model has two $U(1)$ fieldstrength supermultiplets $\Upsilon_{-a}$, under which six chiral multiplets, $\Phi_{i}$, carry charges 
\be
Q^{a}_{i}=\begin{pmatrix}-3&0&0&1&1&~~1\\ ~~0&1&1&0&0&-2\end{pmatrix},
\ee
five left-moving fermi multiplets, $\G_{l}$, carry charges 
\be
q^{a}_{l}=\begin{pmatrix}-3&0&0&1&1\\ ~~0&1&1&0&0\end{pmatrix},
\ee
and one Torsion supermultiplet $\T$ carries complex charge 
\be
N^{a}=N^{a}_{1}+iN^{a}_{2}=(1+i)\begin{pmatrix}1\\ -2\end{pmatrix}.
\ee
These charge assignments are chosen to ensure Green-Schwarz cancellation of the gauge anomaly, 
\be
\sum_{i}Q^{a}_{i}Q^{b}_{i}-\sum_{l}q^{a}_{l}q^{b}_{l}= \half\sum_{A}N^{a}_{A}N^{b}_{A}
\ee
Finally, the superpotential takes the form, 
\be
\W = \G_{0}G(\Phi)+\G_{l}\Phi_{0}J^{l}(\Phi) + (r^{a} + N^{a}\T)\Upsilon_{-a}.
\ee

Ignoring the torsion multiplet, the geometry of the semi-classical Higgs branch at $r^{a}\gg 1$ is a $K3$ hypersurface $G=0$ in $\tilde{W\IP}_{1,1,2,2}$ equipped with a holomorphic vector bundle $\V$ which is inequivalent to $T_{K3}$ over the blown-up $\IP^{1}$, leading to a 1-loop gauge anomaly measuring $ch_{2}(T_{K3})-ch_{2}(\V)$.  This anomaly is precisely cancelled by the classical gauge variation of the axionic coupling $\Upsilon_{-}\T$.  The result is a \nK\ $T^{2}$-fibration over $K3$ with $H$-flux supported over the exceptional divisor.

		\subsubsection{Non-SUSY Models}

Of the many ways to break spacetime \susy\ discussed above, compactification on $T^{4}$ is perhaps the most entertaining.  To build a linear model for this geometry, we'll realize $T^{4}$ as a product of elliptic curves cut out of $\IP^{2}$ by cubic superpotentials.  The GLSM is thus a product of two sub-GLSMs, each a $U(1)$ theory with four scalars of charges
\be
Q_{i}=\begin{pmatrix}-3&1&1&1\end{pmatrix},
\ee
and 4 fermi multiplets of charges,
\be
q_{l}=\begin{pmatrix}-3&1&1&0\end{pmatrix},
\ee
interacting via a cubic superpotential,
\be
\W = \G_{0}G(\Phi)+\G_{l}\Phi_{0}J^{l}(\Phi),
\ee
as usual.  Forgetiing about the anomaly, two copies of this should give us our $T^{4}$; the bundle, however, does not satisfy the anomaly condition, so we must add an axion with complex charge
\be
N^{a}=N^{a}_{1}+iN^{a}_{2}=\begin{pmatrix}1\\ i\end{pmatrix},
\ee
where the radius of $\vt_{l}$ are $\sqrt{2}$ times the usual radius, so as to enforce Green-Schwarz cancellation of the gauge anomaly.    The result, as usual, is a \nK\ $T^{2}$-fibration over $T^{4}$ with $dH\neq0$.  To see whether the spectrum is \susic\ directly in the worldsheet requires checking whether the R-charge is integer-quantized, as required by spacetime \susy.

\section{Intrinsic Torsion} \scz  

In the last section we learned that the target space of a GLSM with Green-Schwarz anomaly cancellation is a \nK\  fibration $T^{2}\to X\to M$ over  a \Ka\ base $M$ with 3-form flux $H$ and gauge flux  $F$ satisfying the Bianchi identity $dH=tr R\wedge R - Tr F\wedge F$.  Importantly, our analysis obtained in the ``large radius'' NLSM regime, $r\gg 1$, of our original GLSM for $M$.  Along the way, the FI parameter, $r$, transmuted from the \Ka\ modulus of a $\IP^{1}$ in $M$ to the squashing parameter of an $S^{3}$ in the total space $X$.  

Notably, the addition of the axion doublet did not modify the running of the FI parameter: thanks to \ZT\ non-renormalization theorems, $r$ remained an exact modulus.  Meanwhile, the axions ensured that the vector was always massive, lifting any would-be classical Coulomb branch and removing the singularity at $r=0$ on the Higgs branch where the Coulomb branch would have appeared.  It thus makes sense to ask what happens as we take $r\to-\infty$.  Classically, this limit corresponds to blowing down the corresponding $\IP^{1}$ in $M$.  What is the the correct quantum description in $X$, and what happens to the CFT?

First, recall what happens in the \Ka\ case (see Sec \ref{OrbGeo}).  When $r\ll-1$, vanishing of the $D$-term potential requires some field with negative charge (let's call this field $p$, with charge $-n$) to acquire a non-zero vev of order $\sqrt{r}$, leaving the rest of the scalars, $\phi_{i}$, massless and interacting only through the superpotential, $W\sim\sqrt{r}G(\phi)$.  
The IR limit is thus controlled by a Landau-Ginsburg model with superpotential $W$.  (If the superpotential is identically zero, $G=0$, this is just the trivial sigma model on $\IC^{d}$.)  
Meanwhile, the higgsing vev leaves a $\IZ_{n}$ subgroup of the gauge group unbroken; quotienting by this discrete gauge group generates a $\IZ_{n}$ orbifold of the LG model (or $\IC^{d}/\IZ_{n}$ if $G=0$).  For $r\ll-1$, then, the theory flows in the IR to an LG orbifold controlled by the UV superpotential.

Now consider the \nK\ case, 
focusing for the moment on Abelian gauge groups (we'll return to the more general case in Sec 5.4). 
When $r\ll-1$, vanishing of the $D$-term again forces a field with charge $-n$ to take a vev, higgsing the gauge group to a discrete subgroup and resulting in a Landau-Ginzburg orbifold.  Since this $\IZ_{n}$ is now a subgroup of an anomalous gauge group, its action on the fields of the LG model is generically also anomalous, leading to a discrete chiral anomaly.  By construction, however, this discrete anomaly is precisely cancelled by the asymmetric $\IZ_{n}$ action on the WZW model.  The net result is a non-anomalous asymmetric orbifold of the combined LG+WZW model.

As we shall see, this cancellation can be nicely recast in terms of a set of sector-dependent phases in the orbifold partition function.  Briefly, this works as follows. The $U(1)$ anomaly evaluated on the generator of the unbroken $\IZ_{n}$ gauge group corresponds to a phase of the form, $\w^{\A h\over n}$, where $\w$ is an $n^{th}$-root of unity and $h$ labels the twist sector.  This phase is then cancelled by a compensating phase from the asymmetric gauging of the WZW sector.  We can thus trade the asymmetric action on the WZW model for the addition, by hand, of these compensating phases directly in the partition function.   Since these phases arise as the blow-down of the intrinsic torsion of the $r\to+\infty$ ``geometry'', and act like intrinsic cousins of the more familiar, and by contrast entirely optional, discrete torsion, it seems reasonable to refer to them as {\em intrinsic torsion} phases.

The remainder of this section is devoted to a derivation of these intrinsic torsion phases by chasing our torsion linear models to orbifold points of the base, and to clarifying the relationship of these intrinsic torsion phases to discrete torsion -- which we will find also corresponds to the blow down of geometric torsion on a smooth geometry.  Once the basic plot is clear, we'll look at a few examples.  

	\subsection{The Discrete Anomaly}

Let's focus for the moment on our canonical toy example, $M=\O_{\IP^{1}}(-2)$; generalizing this will be entirely straightforward.  As we take $r\to-\infty$ at a fixed energy scale, minimizing the D-term potential
$$
U = e^{2}(|z_{1}|^{2}+|z_{2}|^{2}-2|p|^{2}-r)^{2}.
$$
forces $p$ to take a non-zero vev,
$$
|p|^{2}={|r|\over 2},
$$
which higgses the $U(1)$ to the $\IZ_{2}$ subgroup generated by $\a_{*}=\pi$.  Under a gauge transformation by $\a_{*}$, the measure transforms as
$$
\int\!\CD\Phi ~ \stackrel{\a_{*}}{\longrightarrow} \int\!\CD\Phi ~  e^{i\a_{*}\A \int\! {F\over2\pi}} =  \int\!\CD\Phi ~  e^{i\pi\A \int\! {F\over2\pi}} \,,
$$
while the action transforms (thanks to the asymmetric gauging of the WZW model) as,
$$
{1\over4\pi} S~ \stackrel{\a_{*}}{\longrightarrow} {1\over4\pi}S -i\pi\A\int\!\ {F\over2\pi}\,.
$$
So while this $\IZ_{2}$ suffers a quantum anomaly proportional to $\int\!{F\over2\pi}$, the anomaly is again cancelled by the classical anomaly of the WZW model, leaving the full partition sum $\IZ_{2}$-invariant.

Generalizing this to more general models is straightforward.  Suppose our model of choice has a phase $r\ll-1$ in which the role of the higgs is played by a field $p$ of charge $-n$.  Around the minimum of its potential, the gauge group is broken to a $\IZ_{n}$ subgroup which inherits from its parent $U(1)$ a discrete anomaly of the measure, as well as a discrete variation of the classical action which precisely cancels the quantum anomaly.   Green-Schwarz cancellation of the $U(1)$ anomaly ensures that the discrete anomaly cancels, too.

	\subsection{Instantons and the Orbifold Partition Sum}

It is useful to understand exactly how this orbifold action arises.  Consider again the $\IZ_{n}$ model discussed in the paragraph above.   This theory looks well-defined around $r\to-\infty$, so it is reasonable to look for a simple description in the IR.  We'll derive the IR physics by computing an effective description at an intermediate scale near $e^{2}$, then run this effective theory to the deep IR.  Perturbatively, it is easy to check that there is no modification of the effective potential from integrating out the massive vector and higgs field.  However, and rather crucially, there are worldsheet instanton corrections that we must take into account.


Since we are interested in the groundstates of the gauge theory, it suffices to study BPS instantons, ie solutions of the BPS equations,
\bea
\CDB p = 0 \\
F = e^{2}(-n|p|^{2}-r).
\eea
These equations have two types of solutions: constant solutions with $F=0$ and $|p|^{2}=-r/n$; and instanton solutions in which the phase of $p$ winds non-trivially around points in the worldsheet where its norm $|p|$ vanishes, forcing the vector to have non-trivial gradients, $F=-e^{2}r$.  Since our higgs field $p$ has non-minimal charge $-n$, however, shifting the phase of $p$ by a single period $p\to e^{2\pi i}p$ corresponds to a gauge rotation by $-\frac{2\pi}{n}$, which is non-trivial as it lies in the $\IZ_{n}$ gauge group left unbroken by the zero-mode vev of $p$.  The instantons we are actually interested in, then, are fractional anti-instantons with
$$
\int \! {F\over2\pi} = -\frac{h}{n} ~~~~ h>0,
$$
ie solutions in which $p$ vanishes at $h$ generically distinct points, $x_{a}$, on the worldsheet, with the phase of $p$  winding by $2\pi$ (corresponding to a gauge shift of $-\frac{2\pi}{n}$) around each $x_{a}$.  

Since the vector and higgs field are both massive, with $m^{2}_{p}\sim m^{2}_{A}\sim e^{2}r$, all the field  gradients of the instanton background must be concentrated within a region of order $1/e\sqrt{r}$ around the zeroes of the higgs field; outside this region, the higgs field settles into its minimum at $|p|^{2}=-r/n$.    As we flow into the IR, where $e\to\infty$, each zero becomes a point-like instanton of instanton number $-1/n$; as we circle one of the $x_{a}$, all the massless fields transform under an element of the unbroken gauge $\IZ_{n}$ subgroup,
$$
\phi_{i} \to \w^{Q_{i}}\phi_{i}, ~~~~~~ \t_{l}\to\t_{l}-\frac{2\pi}{n}N_{l},
$$
where $\w^{n}=1$.  The zeroes of a given instanton configuration, $x_{a}$, thus act as twist field insertions fuzzed out on the compton scale $1/e\sqrt{r}$.  Deep in the IR, where $e^{2}\to\infty$, summing over instanton configurations in the partition function thus reduces to summing over twist-field insertions.  Away from these insertions, integrating out the massive vector and higgs field leaves us with a free action for the remaining fields.  Summing over insertions then generates a $\IZ_{n}$ orbifold of this free theory.  

However, in the presence of GS anomaly cancellation, this is not an ordinary orbifold.  In particular, the orbifold group acts non-trivially on the (classically anomalous)  action.  In a background with instanton number (aka twist) ${\int \!{F\over2\pi} = -\frac{h}{n}}$, the action transforms under the $g^{th}$ element of $\IZ_{n}$ (with $\a_{g}=\frac{2\pi}{n}g$, $g\in\{0..n\}$) as,
\be\label{eqPhase}
{1\over2\pi}\delta_{g}S_{h} = i\A \int\! \a_{g} {F_{h}\over2\pi} ~=~ -i\A~\!\frac{2\pi g}{n}~\!\frac{h}{n} ~=~ -2\pi i\A~\!\frac{gh}{n^{2}} ~.
\ee
What does this do to the orbifold?  To clarify, let's forget about the variation of the action for the moment and build the 1-loop partition function for the orbifold as usual,
$$
Z = \sum_{[g,h]\in\G} ~ \ZZ{g}{h}   ~ = ~  \sum_{h} ~\lp\sum_{g} g\rp \ZZ{}{h}
$$
where $\sum g \equiv\IP_{\G}$ is the projection operator onto states invariant under the orbifold group $\G$, and 
\be
\ZZ{}{h} = \int_{\rm h-twisted}\hspace{-1cm}{\cal D}\Phi ~~ e^{-S_{free}} 
\ee
is the partition sum restricted to the $h$-twisted sector, ie with the sum over field configurations restricted to those satisfying the periodicity conditions $\Phi(\s+1)=h\Phi(\s)$.  (The box notation here represents the genus-one worldsheet, with the labels specifying the periodicity along each leg of the torus.)  The sum over $h$-twisted sectors is required to ensure modular invariance as follows.  In the untwisted sector, acting with $g$ corresponds to requiring the fields to satisfy 
$$
\Phi(\sigma,\tau+1) = g\Phi(\sigma,\tau), ~~~~\Phi(\sigma+1,\tau) = \Phi(\sigma,\tau),
$$
Under a modular transformation $(\sigma,\tau)\stackrel{SL(2,\IZ)}{\longrightarrow}(a\sigma +b\tau, c\sigma+d\tau)$, this becomes,
$$
\Phi(\sigma,\tau+1) = g^{d}\Phi(\sigma,\tau), ~~~~\Phi(\sigma+1,\tau) = g^{-b}\Phi(\sigma,\tau),
$$
corresponding to a state with $g^{-b}$-twisted boundary conditions.  It is thus impossible to construct a modular invariant orbifold partition sum without including twisted sector states satisfying boundary conditions $\Phi(\s+1,\tau)=h\Phi(\s,\tau)$.  To get a modular invariant path integral, we must sum over $(g,h)$-twisted sectors satisfying the boundary conditions
$$
gh\Phi(\sigma,\tau) = g\Phi(\sigma+1,\tau)=\Phi(\sigma+1,\tau+1),
$$
for all commuting pairs $[g,h]=0\in \G$.  Under a modular transformation $(g,h)\to(g^{d}h^{-c},g^{-b}h^{a})$, the resulting partition sum transforms as,
$$
Z = \sum_{[g,h]\in\G} ~ \ZZ{g}{h}   ~ \to ~ \sum_{[g,h]\in\G} ~ \ZZ{g^{d}h^{-c}}{g^{-b}h^{a}} = Z
$$
where the last step involved relabeling the generators of the orbifold group.

In our flux compactification, however, the anomaly of the measure explicitly spoils modular invariance.  Recall that the anomaly of the measure, $\CD\Phi$, is 
\bea
{\cal D}\Phi ~\!  &\rightarrow&  {\cal D}\Phi ~ e^{\A\!\int \!\alpha {F\over2\pi}} \\
&=& {\cal D}\Phi ~ e^{\frac{2\pi i}{n^{2}}\A gh},
\eea
where $\a = 2\pi i\frac{g}{n}$ and $\int \! {F\over2\pi} = \frac{h}{n}$.
Acting on the twist-$h$ partition sum with $g\in\IZ_{n}$ thus generates additional phases in the partition sum of the form,
\be
g\lp\ZZ{}{h}\rp = \int_{\rm h-twisted}\hspace{-1cm}{\cal D}\Phi ~~ e^{\frac{2\pi i}{n^{2}}\A gh} ~ g\cdot e^{-S_{free}} 
\ee
Under a modular transformation, the full partition function thus transforms as,
\bea
Z &=& \sum_{[g,h]\in\G} ~ \ZZ{g}{h}   \\
  &\to& ~ \sum_{[g,h]\in\G} ~ \ZZ{g^{d}h^{-c}}{g^{-b}h^{a}} \\ 
  &=&  \sum_{[g',h']\in\G} ~  \eps_{\A}({g'}^{d-b},{h'}^{a-c}) ~ \ZZ{g'}{h'},
\eea
where
$$
\eps_{\A}(g,h)=e^{\frac{2\pi i}{n^{2}}\A gh} .
$$
Unless the anomaly vanishes, $\A=0$, the phases $\eps_{\A}$ appear to be an explicit, and disastrous, violation of modular invariance.

This is where the classical $\IZ_{n}$-variation of the action comes in to save the day.  In the above, we were explicitly using the free action, $S_{free}$, in computing the partition sum.  But this is obviously wrong -- had we used the free action in the UV, the anomaly would also have bit us in the neck.  To cancel the anomaly, we added the classically-anomalous axial coupling.  In the IR of the orbifold phase, $r\ll-1$, the vector has been integrated out; the only remaining effect of the axial coupling is thus locked away at the zeroes of the fractional instantons, circling which shifts the full action by an anomaly cancelling phase.  In the orbifold limit, then, we can still use the free action, but {\em only} locally; as we circle a twist field, we must add the phase which derives from the axial coupling.  Since this is a constant phase depending only on the twists $(g,h)$, we can pull it out of the path integral, giving a modified partition sum
\be\label{eqITPhase}
Z = \sum_{[g,h]} ~\eps(g,h) ~  \ZZ{g}{h}  ~~~~~~ \eps(g,h)=e^{-\frac{2\pi i}{n^{2}}\A gh}.
\ee
Under a modular transformation, the classical phase  $\eps(g,h)$ precisely cancels the anomalous phase coming from the measure.  This is nothing but the orbifold limit of the GS mechanism we used in the $r\gg1$ phase, with all the metric, gauge and 3-form flux locked away at the core of the shrunken instantons.

Computationally, it is useful to recast these phases as modifications of the orbifold projection.  In particular, rather than the usual projection onto invariant operators in each $h$-twisted sector, $g\O_{h}=\O_{h}$, our GS orbifold enforces a modified projection,
\be
g \cdot \O_{h} = \eps(g,h)~\!\O_{h}.
\ee
Note that $\eps(0,h)=\eps(g,0)=1$, so the untwisted sector projection is not modified -- the basic orbifold action is unchanged.  However, in an $h$-twisted sector, the orbifold projection now requires not that the ground state is $\G$-invariant, but rather that its $\G$-variation is correlated with the twist sector so as to cancel the anomaly.

These intrinsic torsion phases, and the modified orbifold projection which is their heir, are reminiscent of discrete torsion \cite{Vafa:1986wx}.  In the case of discrete torsion, one begins with a modular invariant orbifold partition function and asks whether it is possible to introduce additional sector-dependent phases to the partition sum while preserving modular invariance so as to give a new orbifold CFT which differs from the original only in the twisted sectors.  As Vafa showed, this is indeed possible, but places tight restrictions on the orbifold group -- in particular, the possible phases are constrained by 1-loop modular invariance and factorization of higher-loop partition sums to represent 2-cocycles of the orbifold group,
$$
\eps(g,h)_{DT} \in H^{2}(\G,\IZ).
$$
%
Notably, such phases may be interpreted as a non-trivial choice of orbifold Wilson-surface for the $B$-field, ie a $B$-field which is flat, $dB=0$, everywhere away from the orbifold fixed points. (This is just a 2-form analog of an orbifold wilson line, a non-trivial action on a 1-form $A$ which leaves $dA=0$ away from fixed points).  The lagrangian term,
$$
\L = \eps^{ab}\p_{a}X^{\mu}\p_{b}X^{\nu} B_{\mu\nu},
$$
evaluated on this orbifold wilson surface then generates the discrete torsion phases $\eps_{DT}$

In the case with {\em intrinsic} torsion, the phases again derive from a non-trivial orbifold wilson surface for $B$.  This time they are not optional but are required to cancel the anomaly.  More explicitly, in the orbifold limit, all the metric and bundle curvature are locked away at the orbifold fixed points.  In particular, the anomaly is supported at the fixed points only.  To cancel this anomaly via Green-Schwarz, we need $B$-field flux locked at the same singularities.  Hence the orbifold $B$-field.  However, unlike the more familiar case of discrete torsion trapped at singularities of type II orbifolds, in this heterotic context we can resolve the geometric singularity by taking $r$ positive.  The resulting smooth geometry supports smoothly varying metric curvature, $R$, bundle curvature, $F$, and non-vanishing $H$-flux related to each other by the GS mechanism (something we could not arrange while preserving the gauged $(1,1)$-\susy\ of the type II worldsheet, which is one way to understand why it is hard to resolve orbifolds supporting discrete torsion in type II string theory).

\subsection{A Compact Example: $T^{2}\to K3$}
\newcommand{\tG}{\tilde{\Gamma}}
Note that all of the above obtains in the presence of a classical superpotential.  Consider a compact example given by a $T^{2}$-fibration over the quartic $K3$ with anomalous bundle pulled back from $\IP^{3}$.  The fields of the associated TLSM and their gauge and $U(1)_{L,R}$ charges are
\be
{\setlength\arraycolsep{8pt}
\begin{array}{|c||c|c|c|c|c|}
\hline  & \boldsymbol{\Phi^{i=1,\ldots,4}}  & \boldsymbol{P} & \boldsymbol{\tG} & \boldsymbol{\G^{m=1,\ldots,5}} & \boldsymbol{\Theta}  \\
\hline\hline \boldsymbol{U(1)}   &  1  &  -5  &  -4  &  1  &  (3R+i S)      \\
\hline \boldsymbol{U(1)_L}   &  \frac{1}{5}  &  0  &  -\frac{4}{5}  &  -\frac{4}{5}  &  \big(  \frac{3}{5}R + \frac{i}{5}S \big)    \\
\hline \boldsymbol{U(1)_R}   &  \frac{1}{5}  &  0  &  \frac{1}{5}  &  \frac{1}{5}  &  \big(  \frac{3}{5} R + \frac{i}{5} S \big)  \\
\hline
\end{array}}
\ee
Here, $R = \frac{1}{\sqrt{3}}$ and $S=1$ are the radii of the $S^{1}$ fibres.  The given charges ensure that all anomalies vanish with $c_L = \hat{c}_R + r_L = 6 + 4$, assuming the superpotential to be the most general function compatible with these charge assignments.

For $r^{a}<< -1$, the gauge groups is higgsed to a $\IZ_{5}$ subgroup, lifting $P$ and leaving the $\Phi_{i}$ massless.
The superpotential hasn't gone anywhere, so the resulting theory is a \ZT\ Landau Ginzburg orbifold.
As before, the orbifold action is the non-anomalous combination of the inherited action on the LG model and the asymmetric action on the axions; equivalently, it is the symmetric action on the axions augmented by intrinsic torsion phases.  As a non-anomalous orbifold of a well-defined CFT, the computational tools developed in \cite{Kachru:1993pg,Distler:1994hs,Berglund:1995dv,Silverstein:1994ih} may be thus applied to compute the massless spectrum.  The explicit computation of the spectrum of this model is presented in \cite{Adams:2009SPEC}.

	\subsection{Fibred WZW Models and Heterotic Non-Geometry}
The WZW presentation of these models (see Sec 4.2) provides a particularly useful way to organize these orbifolds.  In these models, the anomaly of the GLSM sector is cancelled by the classical anomaly of an asymmetrically gauged WZW sector.  The classical phases in the partition function are simply the difference between the symmetric and asymmetric actions evaluated on the discrete orbifold gauge action and in a given topological sector.  This classical variation of the WZW action then cancels the quantum anomaly of the LG model -- as it was constructed to do.

This suggests a structure for more general heterotic flux vacua, and a principle with which to construct new examples.   Start with a classical \ZT\ Landau-Ginzburg model and orbifold by some discrete classical symmetry.  In general, the orbifold action will be anomalous.  To cancel this anomaly, find a \ZT\ WZW model whose lagrangian varies under some choice of action of the orbifold group so as to cancel the one-loop anomaly of the LG model.  Orbifolding the product of the LG and WZW models thus gives a consistent theory with non-trivial Green-Schwarz anomaly cancellation.  The GLSM thus provides a powerful guide to identify satisfactory models, and locates them as special non-geometric points deep inside the moduli space of \nK\ SU(3) manifolds.

Note that this structure -- a (0,2) gauged LG orbifold whose anomaly is cancelled by inflow from a gauged WZW model -- is precisely the form of the (0,2) models studied in \cite{Distler:1993mk}, suggesting that the constructions of \cite{Distler:1993mk} {\em do} have geometric limits as \nK\ torsion compactifications.  In principle, mapping the defining data of those models to a specific spacetime geometry should be a simple matter of constructing the appropriate linear model.  
It would be interesting to flesh out this geometry.

\section{Conclusions and Open Problems}\scz

In this paper, we have used gauged linear sigma models with chiral \susy\ and Green-Schwarz anomaly cancellation to argue that certain flux compactifications of the heterotic string may be blown down to singular limits described by asymmetric LG orbifolds.  Conversely, this paper argues that orbifold CFTs enjoying discrete Green-Schwarz anomaly cancellation may contain twisted moduli which blow-up the orbifold to produce smooth heterotic flux compactifications.  Along the way, we have explained how this generalized CY-LG correspondence allows an exact computation of the spectrum and \susic\ interactions of the full worldsheet theory without any appeal to SUGRA perturbation theory.  Many of these models -- for example, the compact $T^{2}$-fibrations over $K3$ -- are dual to more familiar and less controlled type II orientifolds of smooth Calabi-Yau's supporting non-trivial RR flux.  These models thus provide orbifold CFT descriptions of the dual RR flux-vacua.

A great deal remains to be explored.  The orbifold description allows us to compute the exact spectrum of a generic compact heterotic flux compactification, as well as its quantum cohomology ring.  The computation of such a spectrum will presented in \cite{Adams:2009SPEC}.  One surprise in these constructions is that the supergravity turned out to be as reliable as it did -- is there an all-orders argument for the perturbative existence of these vacua analogous to those which apply to \Ka\ $(0,2)$ compactifications\footnote{Thanks to Shamit Kachru for discussion on this point.}?  Certainly in cases with $\N=2$ spacetime \susy\, and thus $(0,4)$ worldsheet \susy, one can hope for considerable power in constraining corrections to the spacetime superpotential. 

It seems likely that these tools will shed additional light on many other questions surrounding these torsion compactifications.  For example, what is the mirror of a heterotic flux compactification, and what is the global geometry of its moduli space?   Is there any useful relationship between the LG models studied in this paper and the flux-stabilized IIB LG models of \cite{Becker:2006ks}?  Can we build controlled examples with non-CY bases, for example over del Pezzo surfaces, and compute their spectra?  What about models whose $K3$ bases are branched covers of del Pezzo surfaces -- is the branched cover structure manifested in the linear or orbifold descriptions\footnote{I thank Li-Sheng Tseng and collaborators for raising this question.}? How do we build torsion compactifications which are {\em not} elliptic fibrations?  What is the structure of the full landscape of heterotic flux vacua -- in particular, what are some examples which lift all moduli?

\vskip0.3in	
\begin{center}{\bf Acknowledgements} \end{center}
\vskip0.1in
\noindent
I would like to thank 
K.~Becker,
M.~Becker,
S.~Kachru,
A.~Lawrence,
J.~\mbox{McGreevy}.
E.~Silverstein,
W.~Taylor,
L.-S.~Tseng,
S.-T.~Yau
and 
B.~Zwiebach 
for many fun and enlightening conversations, and especially J.~Lapan for collaboration on related projects and countless illuminating discussions.  
Thanks also to the organizers and participants of the Amsterdam Summer String Workshops,  
the Kavli Institute for Theoretical Physics at UCSB 
and the Aspen Center for Physics,
where this work was discussed and completed.  This work was supported in part by the DOE under contract No.~DE-FC02-94ER40818.

\vskip 0.5cm

~
\begin{appendix}

\section{More General Examples}

In this appendix we briefly discuss a few ways one can try to generalize the construction of  \cite{Dasgupta:1999ss,Becker:2006et}.

	\subsection{Models without SUSY}	

More general solutions may be constructed if we relax the condition of spacetime supersymmetry.  The resulting compactifications will generically not satisfy the BPS conditions given above, but only the full equations of motion.  Rather than attempt to build full solutions, we will simply give an ansatz for all the low-energy heterotic fields.  

The most obvious generalization is to take the curvature of the $T^{2}$ bundle to be a general $(1,1)$-form, rather than strictly anti-self-dual (1,1)-form, as was required for supersymmetry.  
But adding a self-dual component to the curvature leads to sign-indefinite terms in the integrability condition deriving from the Bianchi identity, dramatically changing the character of the resulting solutions.  Similarly, we might take the curvature of the vector bundle to be imprimitive, or have $c_{1}(\V_{S})\neq 0$; this is a little more delicate to work with, so we'll stick to tweaking the $T^{2}$-bundle for now.

To get a sense for the doors this opens up, consider the case of the Iwasawa manifold, a $T^{2}$-fibration over $T^{4}$.  In \cite{Becker:2006et}, it was argued that this cannot be a solution of the heterotic BPS equation.  The argument boils down to studying the integrability condition for the Bianchi identity; since $c_{2}(T_{T^{4}})=0$, the integrability condition simplifies to,
$$
\int_{X} e^{4\phi}(||\w^{1,1}_{A}||^{2}-||\w^{1,1}_{S}||^{2})\wedge J^{3} = \int_{X} tr \F\wedge\F\wedge J.
$$
Since the gauge field strength $\F$ is an anti-self-dual anti-hermitian (1,1) form, the right-hand side is non-positive.  Meanwhile, SUSY requires the self-dual part of the (1,1) form to vanish, so the left hand side is non-negative.  The only solution is thus the trivial solution.  If, however, we break SUSY by adding a self-dual (1,1)-piece, $w^{1,1}_{S}\neq 0$, the left hand side is no longer sign-definite and we can find a solution to the integrability condition.  This will not lead to a solution of the BPS conditions, of course; nonetheless, as we shall see via the linear model in the next section, it does lead to a perfectly good \ZT\ worldsheet scft.

Meanwhile, by adding a sign-indefinite term to the integrability condition, breaking susy allows something we previously could not find -- a true large-flux limit.  Let's go back to the integrability condition for the $K3$ example,
$$ 
24-c_{2}(\V_{K3}) = \sum_{i,a}N_{i}^{a}N^{b}_{i} C_{ab} 
$$
Previously, enforcing SUSY made the RHS positive definite, while the left-hand side was bounded by 24, so only a finite number of integer solutions existed.  If we take $w^{1,1}_{S}\neq 0$, however, the RHS is not positive definite; by taking $w^{1,1}_{S}$ and $c_{2}(\V_{K3})$ both large, then, we can find arbitrarily many solutions of the integrability condition.

To take advantage of this vast embiggening of the space of possibilities, it's useful to include a further generalization.  In principle, we do not need the integers appearing in $H$ and $\w_{i}$ to be the same; more generally, we may take 
$$ 
\w_{i} = M_{i}^{b}F_{b} 
~~~~~~ 
H=N^{ia}(d\theta_{i}+M_{i}^{b}\a_{b})\wedge F_{a}
~~~~~~ 
R^{2}_{i}= N^{a}_{i}/M^{a}_{i} ~~ \forall a ~,
$$ 
where the $F_{a}=d\a_{a}$ are unit elements of the cohomology of the base and the $R_{i}$ are the radii of the $S^{1}$-fibres.   Now let's go back to the $T^{4}$ example and consider the limit $M_{i}=1$.  If we take the $N_{i}$ large, repeating the analysis from above gives a curvature invariant which scales with a negative power of $N$; at large flux-numbers, then, the manifold is weakly coupled and may be reliably described by SUGRA perturbation theory.  Of course, since this compactification breaks SUSY, it is not clear that these geometries make sense quantum-mechanically.  To answer this requires a computation of the spectrum.  

	\subsection{More General Bases}	

Another obvious move is to replace our base $K3$ by another geometry.  We saw a special case of this above -- the Iwasawa compactification on $T^{4}$.  One simple direction is to work with higher-dimensional bases -- for example, it is relatively straightforward to repeat the above analysis over a CY 3-fold base.  However, for phenomenological reasons, this is not an obviously desirable approach. 

Considerably more promising is compactification on \Ka\ 2-folds with positive curvature.  Such constructions have proven extremely useful in the context of CY constructions, where non-trivial elliptic fibrations over $dP_{k}$ lead to many interesting manifolds.  To motivate such heresy in the present \nK\ context, note that a beautiful theorem of Michelsohn \cite{Michelsohn:1982} tells us that any holomorphic $T^{2}$ fibration over a balanced manifold is again balanced.  (Since every balanced 2-fold is \Ka, we loose no generality in working over a \Ka\ base.)  Thus such a construction immediately admits a solution of all the BPS equations except the BI.  Since the base is not CY, however, the existence of a nowhere-vanishing 3-form on the total space necessitates degeneracies in the fibration as studied in \cite{Greene:1989ya}; relaxing \Ka\ to balanced considerably weakens the constraints, however, making the analysis considerably more complicated\footnote{Since certain K3's form branched covers of del Pezzo surfaces, one can construct such compactifications by orbifolding torsional 3-folds built over these K3's \cite{Becker:2008rc}.  Unfortunately, the resulting singularities remain difficult to interpret -- the theory of balanced resolutions is much less developed than its \Ka\ counterpart -- and the general problem remains open.}.

\section{Counting Vacua on the Heterotic Landscape}

The tools presented in this paper allow the construction of a large number of heterotic flux compactifications.  Well, how many?  As usual, a precise counting requires a very detailed analysis, which we will not be able to provide -- in particular, all of the above vacua include at least one surviving modulus (the heterotic dilaton is never lifted at tree level), and frequently many more.  Nonetheless, it is useful to get a sense for the number of inequivalent families.  We'll focus, for simplicity, on $\N=1$ vacua.

So: how many $\N=1$ families can we build over $K3$?  Well, picking a vacuum involves a choice of bundle $\V_{K3}$ which solves the tadpole condition,
$$
M\cdot N= 24-c_{2}(\V_{K3})
$$
We can thus take $c_{2}(\V_{K3})$ to be anything between 24 and 0; for each possible value, we must then find solutions for the $N$.  A choice of $c_{2}(\V_{K3})$ does not, of course, completely specify the bundle -- many other invariants, eg $c_{1}(\V_{K3})$, are {\em a priori} independent, and must be chosen to fully specify the bundle.  

Physically, however, these are {\em not} completely independent invariants.  For example, to ensure a solution to the BPS conditions, the DUY theorem requires our bundle $\V_{K3}$ to be {\em stable}, ie to be such that any sub-bundle $\E\subset\V$ satisfies,
$$
\frac{c_{1}(\E)}{\rk\E} < \frac{c_{1}(\V)}{\rk\V}.
$$
Roughly speaking, this condition means that the bundle cannot fall apart into a sum of sub-bundles\footnote{In general, the difference between a bundle, $\V$, and a sub-bundle, $\E\subset\V$, is a $K$-theory class $(\V,\E)$ which cannot be represented by a bundle.  For our purposes, we can mostly ignore such subtleties.}, ie that its moduli space is locally smooth.  (Since $c_{1}(\V)\equiv\int\!J^{n-1}\wedge {\rm Tr} F_{\V}$, this condition depends sensitively on our location in the \Ka\ cone of the base -- as we move in the \Ka\ cone, we may pass through chamber walls of marginal stability where the bundle goes unstable and decays into a sum of distinct $K$-classes.)
Meanwhile, by a theorem of Bogomolov, any stable holomorphic bundle of rank $r$ satisfies the further inequality,
$$
\lp\frac{2}{r}c_{2}(\V)-\frac{r-1}{r^{2}}[c_{1}(\V)]^{2}\rp J^{n-2}\ge0.
$$
ie$$[c_{1}(\V)]^{2}    \le    \frac{2r}{r-1} c_{2}(\V).$$
For our \susic\ canonical examples, $c_{1}$ takes value in $H^{2}(K3,\IZ)\cap H^{1,1}(K3)$.  For a given value of $c_{2}(\V)$ between 24 and 0, then, $c_{1}$ is restricted to lie within an $h^{1,1}$-dimensional sphere of radius $\sqrt{\frac{2r}{r-1}c_{2}}$ around the origin.  This is very similar to what happens in more familiar KKLT-like vacua.  Summing over all such possibilities gives
$$
N \sim \sum_{k=0}^{c_{2}(T_{K3})=24}\left[\frac{32\pi}{15}k\right]^{11}\frac{1}{11!}(24-k)\pi\sim10^{18}
$$
inequivalent (families of) $\N=1$ vacua.  

This is, of course, a very coarse estimate.  In particular, it is a spectacular undercount: we assumed $\N=1$ susy for absolutely no reason but simplicity.  To get a sense for how the story changes when we break supersymmetry, imagine adding $NS5$- and $\overline{NS5}$-branes.  In their presence, the tadpole condition on  becomes,
$$
dH= c_{2}(T_{X})-c_{2}(\V_{X}) -[NS5] + [\overline{NS5}] ~~~\Rightarrow ~~~ N^{2}_{1}+N^{2}_{2}= 24-c_{2}(\V_{K3}) -[NS5] + [\overline{NS5}]
$$
Adding $\overline{NS5}$-branes thus adds additional positive terms to the RHS, effectively replacing $24$ by $24+[\overline{NS5}]$ in the above counting, making the number of (families of) vacua scale as,
$$
N\sim (24+[\overline{NS5}])^{b^{2}}.
$$

This is only a small corner of the landscape of heterotic flux compactifications.  A thorough exploration will require a detailed understanding of the remaining moduli, and their potentials.  In particular, the dilaton remains completely unconstrained in our analysis.

\end{appendix}


\bibliographystyle{hunsrt}
\bibliography{TORBrefs}

\end{document}